\documentclass[useAMS,usenatbib]{modified_mn2e}
\input{psfig}

\def\xx{$\xi(r_\sigma,r_\pi)$}
\def\xg{$\xi_R(r)$}
\def\wp{$w_p(r_p)$}
\def\hmpc{$h^{-1}$Mpc}
\def\hkpc{$h^{-1}$kpc}
\def\etal{et\,\,al.}

\def\mc{$M_c$}

\def\hmsol{$h^{-1}$M$_\odot$}
\def\kms{km\,s$^{-1}$}
\def\rsig{$r_\sigma$}
\def\rpi{$r_\pi$}

\def\kmsmpc{km\,s$^{-1}$\,Mpc$^{-1}$}
\def\gad{{\scriptsize GADGET}}
\def\cmbfast{{\scriptsize CMBFAST}}
\def\blin{b_{\rm lin}}
\def\bg{b_g}

\def\rvir{R_{\rm vir}}
\def\navg{\langle N\rangle_M}

\def\nsat{\langle N_{\rm sat}\rangle_M}

\def\mmin{M_{\rm min}}

\def\om{\Omega_m}
\def\s8{\sigma_8}
\def\av{\alpha_v}
\def\avc{\alpha_{vc}}
\def\pkzr{P_{0/R}}
\def\pkqp{P_{2/0}}
\def\xizr{\xi_{0/R}}
\def\xiqp{\xi_2/(\xi_0-\bar{\xi}_0)}
\def\xiQp{Q_\xi}

\def\rhalf{r_{\xi/2}}
\def\rhalfa{r_{\xi/2}(0.1)}

\def\ma{[$\Omega_m, \alpha_v$]}
\def\mb{[$\sigma_8, \alpha_v$]}
\def\mc{[$\beta, \alpha_v$]}
\def\md{[$\Omega_m, \sigma_8$]}
\def\plin{P_{\rm lin}(k)}
\def\pkqplin{P_{2/0}^{\rm lin}}
\def\pkzrlin{P_{0/R}^{\rm lin}}

\def\xiqplin{Q_\xi^{\rm lin}}

\def\pzkmu{P_Z(k,\mu)}
\def\bfit{\beta_{\rm fit}}

\def\spplane{$r_\sigma - r_\pi$}

\def\lcdm{$\Lambda$CDM}

\def\one_col_fig{8.0cm}
\def\two_col_fig{12.7cm}

\title[Redshift-Space Distortions with the HOD]
{Redshift-Space Distortions with the Halo Occupation Distribution  I: Numerical Simulations}
\author[Tinker et.\ al.]{Jeremy L. Tinker,$^{1}$\thanks{E-mail:
tinker@astronomy.ohio-state.edu}
David H. Weinberg,$^{1}$
\& Zheng Zheng$^{2,3}$\\ \\
$^{1}$Department of Astronomy, 
The Ohio State University, 140 W. 18th Avenue, Columbus, Ohio 43210 \\
$^{2}$Institute for Advanced Study, School of Natural Sciences, 
Einstein Drive, Princeton, NJ 08540 \\
$^{3}$Hubble Fellow
}

\begin{document}

\date{}

\pubyear{2005}

\maketitle

\begin{abstract}

We show how redshift-space distortions of the galaxy correlation
function or power spectrum can constrain the matter density parameter
$\om$ and the linear matter fluctuation amplitude $\s8$.  We improve
on previous treatments by adopting a fully non-linear description of
galaxy clustering and bias, which allows us to achieve the accuracy
demanded by larger galaxy redshift surveys and to break parameter
degeneracies by combining large-scale and small-scale
distortions. Given an observationally motivated choice of the initial
power spectrum shape, we consider different combinations of $\om$ and
$\s8$ and find paramters of the galaxy halo occupation distribution
(HOD) that yield nearly identical galaxy correlation functions in real
space.  We use these HOD parameters to populate the dark matter halos
of large N-body simulations, from which we measure redshift-space
distortions on small and large scales.  We include a velocity bias
parameter $\alpha_v$ that allows the velocity dispersions of satellite
galaxies in halos to be systematically higher or lower than those of
dark matter.  Large-scale distortions are determined by the parameter
combination $\beta \equiv \om^{0.6}/b_g$, where $b_g$ is the bias
factor defined by the ratio of galaxy and matter correlation
functions, in agreement with the linear theory prediction of parameter
degeneracy.  However, linear theory does not accurately describe the
distortions themselves on scales accessible to our simulations.  We
provide fitting formulas to estimate $\beta$ from measurements of the
redshift-space correlation function or power spectrum, and we show
that these formulas are significantly more accurate than those in the
existing literature.  On small scales, the ``finger-of-god''
distortions at projected separations $\sim 0.1$ \hmpc\ depend on
$\Omega_m\alpha_v^2$ but are independent of $\sigma_8$, while at
intermediate separations they depend on $\sigma_8$ as well.  One can
thus use measurements of redshift-space distortions over a wide range
of scales to separately determine $\Omega_m$, $\sigma_8$, and
$\alpha_v$.

\end{abstract}



\section{Introduction}

In a universe that obeys the cosmological principle, the clustering of
galaxies is statistically isotropic. But in galaxy redshift surveys
the distances to galaxies are inferred from redshifts, making the line
of sight a preferred direction. Peculiar velocities produce anisotropy
in redshift-space clustering on all scales. On small scales, the
random motions of galaxies in virialized systems stretch groups and
clusters into so-called ``fingers-of-god'' (FOG). On large scales,
coherent flows created by gravity compress overdense regions along the
line of sight and stretch underdense regions correspondingly. Small
and large scale distortions provide diagnostics for the matter density
parameter $\om$ and the amplitude of mass fluctuations (Peebles 1976;
Sargent \& Turner 1977; Kaiser 1987). In this paper and its companion,
we develop techniques for modeling redshift-space distortions that
draw on recent developments in the theory of galaxy clustering. These
techniques are designed to reach the level of accuracy demanded by the
new generation of large galaxy redshift surveys, such as the
Two-Degree Field Galaxy Redshift Survey (2dFGRS; \citealt{colless01})
and the Sloan Digital Sky Survey (SDSS; \citealt{york00}).

In the linear theory model of Kaiser (1987), the relation of the
anisotropic, redshift-space galaxy power spectrum $\pzkmu$ to the
isotropic, real-space galaxy power spectrum $P_R(k)$ is

\begin{equation}
\label{e.kaiser}
P_Z(k,\mu) = P_R(k)(1+\beta\mu^2)^2,
\end{equation}

\noindent where $\mu$ is the cosine of the angle between the
wavevector ${\bf k}$, and the line of sight. The amplitude of the
distortion is determined by $\beta=\Omega_m^{0.6}/\blin$, where the
linear bias parameter $\blin\equiv \delta_g/\delta_m$ is assumed to be
independent of scale ($\delta_g$ and $\delta_m$ represent galaxy and
mass density contrasts, respectively). Fourier transformation of
equation (\ref{e.kaiser}) gives expressions for the galaxy correlation
function in redshift space, \xx\ (Hamilton 1992).

Unfortunately, non-linear effects make equation (\ref{e.kaiser})
inaccurate on all scales where observations yield precise measurements
(Cole, Fisher, \& Weinberg 1994). The effects of non-linearity can be
approximated by a phenomenological model in which galaxies have, in
addition to linear theory distortions, random small scale velocities
drawn from an exponential distribution with dispersion $\sigma_v$
(\citealt{pd94}; Park \etal\ 1994; Cole \etal\ 1995). In this model,
the Kaiser formula becomes

\begin{equation}
\label{e.lin_exp}
P_Z(k,\mu) = P_R(k)(1+\beta\mu^2)^2 (1+k^2\sigma_v^2 \mu^2/2)^{-2}.
\end{equation}

\noindent In practice, most estimates of $\beta$ from large-scale
redshift-space distortions have utilized this linear-exponential
model\footnote{There are several minor variants of this model that
have also been utilized, such as replacing the exponential
distribution with a Gaussian (Peacock \& Dodds 1994) or specifying
that the pairwise distribution of galaxy peculiar velocities is
exponential (e.g. Hatton \& Cole 1999).}, expressed in terms of the
power spectrum as in equation (\ref{e.lin_exp}) or in terms of the
correlation function or spherical harmonics. The current
state-of-the-art measurement is the analysis of the 2dFGRS presented
by Hawkins \etal\ (2003), yielding $\beta=0.49 \pm 0.09$, updating the
earlier 2dFGRS analysis of Peacock \etal\ (2001). Previous
observational efforts and theoretical developments are expertly
reviewed by Strauss \& Willick (1995) and Hamilton (1998).

The essential limitation of equation (\ref{e.lin_exp}) is that it is
derived from an unphysical model. There are several sources of
non-linearity in redshift-space distortions in addition to small scale
dispersion (Cole \etal\ 1994; Fisher \& Nusser 1996), and the
dispersion itself is correlated with the local density and is not a
constant for all galaxies. Scoccimarro (2004) shows that the velocity
distribution corresponding to the linear-exponential model is itself
unphysical, containing a $\delta$-function and a discontinuity at the
origin, and that equation (2) does not become fully accurate even at
very large scales. Hatton \& Cole (1999) concluded that this model
introduces a $\sim 15\%$ systematic error in the determination of
$\beta$, which is significant compared to the precision achievable
with 2dFGRS and the SDSS. Furthermore, the $\sigma_v$ parameter, while
related to the amplitude of the small scale distortions, has no
clearly defined physical meaning. In redshift-space distortion
analyses it is purely a nuisance parameter, significantly degenerate
with $\beta$, and has no use in constraining cosmological parameters.

The program initiated by Kaiser (1987) largely supplanted an earlier
tradition of using small-scale redshift distortions to constrain $\om$
via the ``cosmic virial theorem'' (Peebles 1976, 1979; Davis, Geller,
\& Huchra 1978; Bean \etal\ 1983). The analytic expression of this
``theorem'' relied on the assumption of stable clustering, which early
N-body simulations showed was unlikely to hold on the relevant scales
(e.g., \citealt{davis85}). A more serious problem is that the bias
between galaxy and dark matter clustering is likely to have a complex
effect on quantities that enter the cosmic virial theorem, one that
cannot be captured by a single bias parameter with an obvious physical
interpretation.

The goal of this paper and its companion is to present techniques for
physical modeling of redshift-space distortions that can take
advantage of high-precision measurements on large and small scales. We
construct these techniques in the framework of the Halo Occupation
Distribution (HOD; see, e.g. Ma \& Fry 2000; Peacock \& Smith 2000;
Seljak 2000; Benson 2001; Scoccimarro \etal\ 2001; Berlind \& Weinberg
2002; Cooray \& Sheth 2002), in which the bias of a specified class of
galaxies is defined by the probability distribution $P(N|M)$ that a
halo of mass $M$ contains $N$ galaxies, together with prescriptions
for spatial and velocity bias within individual halos. The HOD has
proven to be a powerful tool for encapsulating the bias predictions of
galaxy formation models (Kauffmann \etal\ 1997; Benson \etal\ 2000;
White \etal\ 2001; Yoshikawa \etal\ 2002; Berlind \etal\ 2003;
Kravtsov \etal\ 2004; Zheng \etal\ 2004), for analytic calculations of
galaxy clustering statistics (see Cooray \& Sheth 2002 and numerous
references within), and for empirical modeling of galaxy clustering
data (Jing \etal\ 1998; van den Bosch \etal\ 2003; Zehavi \etal\
2004a,b; Yang \etal\ 2004; Mo \etal\ 2004; \citealt{kev04,tinker04}).
Several recent papers have presented calculations of redshift-space
distortions or peculiar velocity statistics using halo models of dark
matter and galaxy clustering (Seljak 2001; White 2001;
\citealt{Sheth01a,Sheth01b}; Kang \etal\ 2002; Cooray 2004), providing
insight into the role of non-linear dynamics and non-linear bias in
shaping clustering and anisotropy. However, these studies primarily
focus on dark matter rather than galaxy clustering, and they have not
yet yielded a clear blueprint for constraining cosmological parameters
with HOD modeling of observed redshift-space distortions, which is our
objective here.

We use the HOD formulation to set up the redshift-space distortion
problem in the following terms. Any redshift survey large enough to
yield useful measurements of large-scale anisotropy will first allow
precise measurements of the projected correlation function, \wp, which
is unaffected by peculiar velocities. For any choice of cosmological
parameters, one should choose HOD parameters to reproduce this
measurement of real-space clustering. If an acceptable fit cannot be
found for the given cosmology, then the model is already ruled out
(e.g. Abazajian \etal\ 2004). For models with acceptable real-space
clustering, one calculates redshift-space distortions using numerical
simulations or analytic approximations to test the model's
cosmological parameters. In practice, the parameters that enter are
$\om$ and the amplitude of the linear theory matter power spectrum
$\plin$, which we characterize by $\s8$, the rms linear matter
fluctuation in 8 \hmpc\ spheres (with $h\equiv H_0/100$ \kmsmpc ). We
assume that the {\it shape} of $\plin$ is known from measurements of
the large scale galaxy power spectrum and cosmic microwave background
(CMB) anisotropy, which together pin down the parameters that
determine $\plin$ quite accurately (e.g.,
\citealt{percival02,spergel03,tegmark04a}). Since redshift-space
anisotropy is insensitive to the shape of $P_R(k)$ --- in equations
(\ref{e.kaiser}) and (\ref{e.lin_exp}) the $\mu$-dependence of
$P_Z(k)$ factors out entirely --- small uncertainties in the shape of
$\plin$ should have minimal effect. In this work we adopt the power
spectrum form of Efstathiou, Bond, \& White (1992), where the shape is
parameterized by the characteristic wavenumber $\Gamma$.

While matching \wp\ can constrain HOD parameters relevant to
real-space clustering, we must also allow for the possibility that
galaxies in a halo have a systematically different velocity dispersion
from that of the halo dark matter. (The mean velocity of galaxies and
dark matter within a halo should be the same because both components
feel the same large-scale gravitational field.) Numerical simulations
predict that the galaxy closest to the halo center of mass moves at
nearly the center of mass velocity while satellite galaxies have a
velocity dispersion similar to that of the dark matter
(\citealt{berlind03,faltenbacher04}). We define the satellite
``velocity bias'', $\av$, as the ratio between these two
dispersions. Although the numerical simulations predict that $\av
\approx 1$, this parameter could depart modestly from unity as a
result of dynamical friction, tidal disruption or mergers of slowly
moving satellites, or different orbital anisotropy of galaxies and
dark matter. We will treat $\av$ as a free parameter to be constrained
by the observations, but we will assume that it is constant over the
relevant range of halo masses. We will also consider effects of
non-zero velocities for central galaxies, though simulations predict
these velocities to be $\la 20\%$ of the virial velocity.

In this paper we use N-body simulations to create halo populations for
a set of cosmological models, and we populate those halos with galaxies
using HOD models that yield similar real-space clustering. We examine
the constraints that redshift-space distortions can impose within the
three-dimensional parameter space ($\om$, $\s8$, $\av$), and we use
our numerical results to obtain fitting formulas that can estimate
parameters from observational data. In a companion paper, we develop a
numerically calibrated analytic model for redshift-space
distortions. The analytic model provides physical insight into the
numerical results, and it can make more complete use of the
observational measurements for cosmological parameter estimation.

In Section 2 below, we describe the numerical simulations and the HOD
models used to populate them with galaxies. Section 3 presents an
overview of redshift-space anisotropies in the two-dimensional
correlation function \xx. In \S 4 we focus on measures of large-scale
distortion based on multipole decomposition of the power spectrum and
the correlation function. These measures mainly constrain the
parameter combination $\beta\equiv\om^{0.6}/b_g$, which can be related
to $\s8\om^{0.6}$ using the measured (real-space) galaxy
clustering. (As discussed in \S 4.4, we define $b_g$ by a ratio of
non-linear correlation functions, which makes it similar but not
identical to the linear theory bias factor $\blin$.) In \S 5 we turn
to small scale distortions, which most directly constrain $\om \av^2$
and have some power to break degeneracies further and yield separate
determinations of $\om$, $\s8$, and $\av$. In \S 6 we summarize our
results and discuss how they can be applied to cosmological parameter
estimation from observational data.



\begin{table*}
  \caption{Properties of the Simulations and HOD Parameters}
  \begin{tabular}{@{}ccccccc@{}}
  \hline

  $z_{\rm out}$ & $\om$ & $\s8$ & $b_g$ & $\mmin$ [\hmsol ] 
  & $M_1$ [\hmsol ] & $\alpha$ \\

 \hline

0  & 0.100 & 0.950 & 0.922 & 3.73$\times 10^{11}$ & 9.38$\times 10^{12}$ & 0.934 \\
0.19 & 0.158 & 0.900 & 0.956 & 5.96$\times 10^{11}$ & 1.45$\times 10^{13}$ & 0.959 \\
0.56 & 0.297 & 0.801 & 1.041 & 1.09$\times 10^{12}$ & 2.51$\times 10^{13}$ & 1.005 \\
0.97 & 0.459 & 0.699 & 1.181 & 1.70$\times 10^{12}$ & 3.51$\times 10^{13}$ & 1.109 \\
1.45 & 0.620 & 0.599 & 1.358 & 2.19$\times 10^{12}$ & 3.99$\times 10^{13}$ & 1.199 \\

\hline
\end{tabular}

\medskip
Note. --- When we scale an output to a different value of
$\om$, the values of $\mmin$ and $M_1$ scale in proportion to $\om$,
as discussed in \S 2.3. \\

\end{table*}

\section{Numerical Simulations and HOD Models}

\subsection{N-body Simulations}

We use N-body simulations to create halo populations for a sequence of
cosmological models, always assuming a spatially flat universe
dominated by cold dark matter and a cosmological constant (\lcdm ),
with Gaussian initial conditions and a primordial power spectrum
motivated by observations of CMB anisotropies and large-scale
structure.

We choose the mass resolution by requiring that there be at least 30
particles in the lowest mass halos that host simulated galaxies. On
this basis we select a mean interparticle separation of
$\bar{n}^{-1/3}=0.7$ \hmpc\ for all initial conditions. For $\om=0.3$,
the 30-particle limit corresponds to a minimum halo mass of $\sim
10^{12}$ \hmsol, similar to the minimum halo mass found for the HOD
fit (assuming $\om=0.3$) to the SDSS sample of galaxies brighter than
$M_r=-20+5\log h$ (Zehavi \etal\ 2004b). All of our simulated galaxy
populations have a space density of $\bar{n}_g=5.6\times 10^{-3}$
(\hmpc)$^{-3}$, equal to that of SDSS galaxies brighter than
$0.68\,L_\ast$ (\citealt{blanton03}), or $M_r=-20.04+5\log h$.

To cover the $(\om, \s8)$ parameter space in an efficient manner, we
draw on the findings of Zheng \etal\ (2002), who demonstrated that
changes in $\om$ at fixed $\Gamma$ and $\s8$ simply scale halo masses
in proportion to $\om$ and halo velocities in proportion to
$\om^{0.6}$. In terms of these scaled masses and velocities, the mass
function, spatial correlations, and velocity correlations of halos
identified at fixed overdensity are virtually independent of $\om$.
We can therefore run a single simulation that has a high value of
$\s8$ at redshift zero and use the earlier redshift outputs to
represent $z=0$ results for lower values of $\s8$. For each $\s8$, the
halo population can be scaled to any desired value of
$\om$. Specifically, we run simulations with $\om=0.1$ and $\s8=0.95$
at $z$=0, and use the outputs at $z$=0.19, 0.56, 0.97, and 1.45 when
$(\om, \s8)$ = (0.16, 0.90), (0.30, 0.80), (0.46, 0.70), and (0.62,
0.60). We model different values of $\om$ by scaling the halo masses
in proportion to $\om$, the halo velocities by $\om^{0.6}$, and the
internal halo velocity dispersions by $\om^{0.5}$. We carry out a test
of this scaling in \S 2.3 to demonstrate that it is accurate
enough for our purposes here.

\begin{figure}
\centerline{\psfig{figure=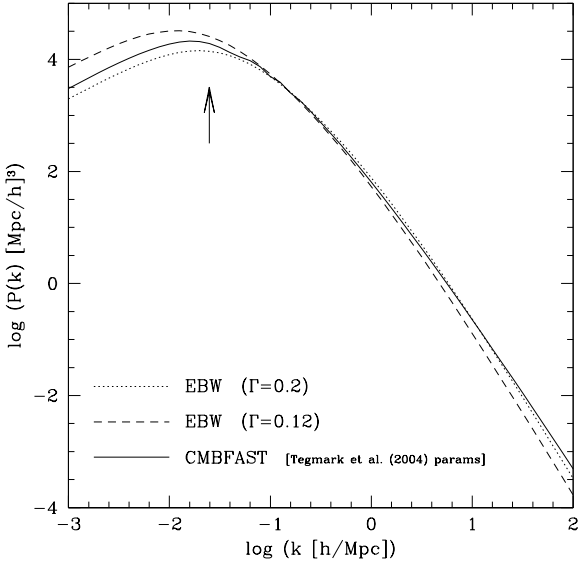,width=\one_col_fig}}
\caption{ \label{psp} The two power spectra used in the simulations,
  $\Gamma=0.2$ and $\Gamma=0.12$, are compared to the linear power
  spectrum computed with \cmbfast\ using the parameters listed in
  column 6 of Table 4 in Tegmark \etal\ (2004b). All three power
  spectra are normalized to the same value of $\s8$. The fundamental
  mode of the simulation volume is indicated with the arrow at
  $k=2\pi/L_{\rm box}=0.025\;h/$Mpc. The $\Gamma=0.2$ power spectrum
  is similar to the \cmbfast\ calculation over the range of scales
  simulated, while the $\Gamma=0.12$ power spectrum has more power at
  large scales and increasingly less as $k$ increases.  }
\end{figure}

We analyze simulations with two values of the power spectrum shape
parameter, $\Gamma=0.2$ and 0.12, both with inflationary spectral
index $n_s=1$.  On the scales probed by our simulations, $\Gamma=0.2$
corresponds well to the power spectrum calculated with \cmbfast\
(\citealt{cmbfast}) with $\om=0.3$, $h=0.7$, and $\Omega_b=0.04$,
values favored by recent observations (e.g., Spergel \etal\ 2003;
\citealt{tegmark04b}). The redder, $\Gamma=0.12$ power spectrum
corresponds to a lower combination of $\om h$, or a tilted ($n_s<1$)
primordial spectrum. This model is at the extreme edge of those
allowed by current data, so comparing results for $\Gamma=0.2$ and
$\Gamma=0.12$ should give a conservative estimate of uncertainties
associated with the power spectrum shape. In Figure \ref{psp} we
compare these two power spectra to one created with the transfer
function calculated by \cmbfast\ using the cosmological parameters
listed in Table 4 (column 6) of Tegmark \etal\ (2004b), who derive
combined constraints from WMAP CMB data, and the SDSS galaxy power
spectrum. Each power spectrum is normalized to the same value of
$\s8$.  The fundamental mode of the box is marked with the
arrow. Inside this scale, the $\Gamma=0.2$ power spectrum closely
tracks the \cmbfast\ calculation. The $\Gamma=0.12$ $P(k)$ has less
small-scale power, but it has significantly more power at scales near
the fundamental mode.

We use the publicly available tree-code \gad\ (Springel, Yoshida, \&
White 2000) to integrate the initial conditions.  We evolve $360^3$
particles in a volume 253 \hmpc\ on a side, giving us a mass
resolution of $9.66\times 10^{10} \times \Omega_m$ \hmsol\ per
particle. The force softening was set to one-tenth the mean
interparticle separation, or $\epsilon = 70$ \hkpc. The simulations
were started at an expansion factor $a=0.01$, with a maximum timestep
of 0.005 in $a$. \gad\ employs individual particle timesteps governed
by a particle's acceleration, such that $\Delta a \propto
\sqrt{\epsilon \eta}$. The value of $\eta$ was set to 0.2. We ran five
independent realizations to estimate the sample variance.

We also ran a similar series of simulations using the particle-mesh
(PM) technique, with a staggered-mesh algorithm similar to that of
\cite{melott83} and \cite{park90}. (The code we use was written by
V. Narayanan.) The high efficiency of the PM algorithm allowed us to
run simulations with the same mass resolution but box sizes of $324$
\hmpc\ per side, twice the volume of our \gad\ runs. In comparing the
results from the two methods, we found that the lower force resolution
of the PM technique (with a $900^3$ grid) had a significant impact on
the number of halos near our 30-particle resolution limit, while the
smaller volume of the \gad\ runs did not adversely affect the
distortions at large scales. We therefore use the \gad\ runs
exclusively in our subsequent analyses.


\begin{table*}
  \caption{Properties of the Mock Galaxy Distributions}
  \begin{tabular}{@{}cccccccccccccccccccc@{}}
  \hline

  \multicolumn{5}{c}{\ma} &
  \multicolumn{5}{c}{\mb} &
  \multicolumn{5}{c}{\mc} &
  \multicolumn{5}{c}{\md} 
  
  \\ \hline

  & $\om$ & $\s8$ & $\beta$ & &
  & $\om$ & $\s8$ & $\beta$ & &
  & $\om$ & $\s8$ & $\beta$ & &
  & $\av$ & $\avc$ & $\beta$ & \\

  \hline

  & 0.3 & 0.95 & 0.53 &  &
  & 0.1 & 0.8 & 0.24 &  &
  & 0.24 & 0.95 & 0.46 &   &
  & 0.0 & 0.0 & 0.46 &  \\

  & 0.3 & 0.90 & 0.51 &  &
  & 0.2 & 0.8 & 0.36 &  &
  & 0.26 & 0.90 & 0.46 &  &
  & 0.8 & 0.0 & 0.46 &  \\

  & 0.3 & 0.80 & 0.46 & &
  & 0.3 & 0.8 & 0.46 &  &
  & 0.3 & 0.80 & 0.46 &  &
  & 1.0 & 0.0 & 0.46 &  \\

  & 0.3 & 0.70 & 0.41 &  &
  & 0.4 & 0.80 & 0.55 &  &
  & 0.36 & 0.70 & 0.46 &  &
  & 1.2 & 0.0 & 0.46 &  \\

  & 0.3 & 0.60 & 0.36 &  &
  & 0.5 & 0.80 & 0.63 &  &
  & 0.47 & 0.60 & 0.46 &  &
  & 1.0 & 0.2 & 0.46 &  \\

  \hline
  \end{tabular}

\medskip
Note. --- In the first three sequences, $\av=1$ and
  $\avc=0$. The HOD parameters and bias factors $\bg$ for each value of
  $\s8$ are listed in Table 1.
\end{table*}

\subsection{HOD Models}

To identify halos in the dark matter distribution we use the
friends-of-friends algorithm (Davis \etal\ 1985) with a linking length
of 0.2 times the mean interparticle separation. Objects identified
with this linking length typically have an average density of
$\rho/\bar{\rho} \sim 200$, which is roughly the criterion for
virialization of a collapsed object. Only halos with 30 or more
particles were retained in the halo sample.

We need to populate the halos with galaxies in a way that generates
similar \xg\ for all values of $\s8$. We use the HOD parameterization
of Kravtsov \etal\ (2004) and Zheng \etal\ (2004), which was also
adopted in the empirical modeling of the SDSS correlation function by
Zehavi \etal\ (2004b). Halos above a minimum mass $\mmin$ are assigned
one central galaxy. The mean number of satellite galaxies in halos
with $M\ge \mmin$ is

\begin{equation}
\nsat = \left( \frac{M}{M_1} \right)^{\alpha}.
\end{equation}

\noindent The mean number of galaxies in a halo is therefore $\navg =
1+(M/M_1)^\alpha$ for $M \ge \mmin$ and $\navg=0$ for $M < \mmin$. We
assume Poisson scatter in the number of satellite galaxies with
respect to the mean $\nsat$, consistent with the theoretical
predictions of \cite{kravtsov04}, and \cite{zheng04}.

\begin{figure}
\centerline{\psfig{figure=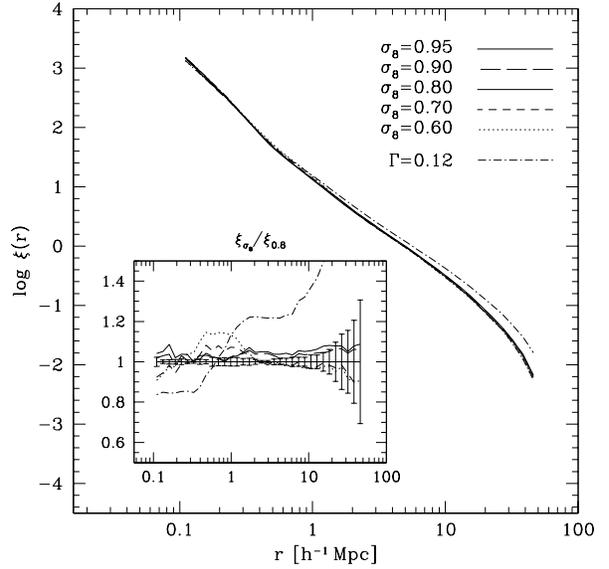,width=\one_col_fig}}
\caption{ \label{real_space_xi} The real-space galaxy two-point
  correlation functions for the five cosmologies and HOD parameters
  listed in Table 1. The inset box shows the different correlation
  functions normalized by that of the central model, $\om = 0.3$, $\s8
  = 0.8$. The error bars in the inset box are those for the central
  model. Results are averaged over five realizations, and error bars
  show the run-to-run dispersion divided by $\sqrt{N-1}=2$ to
  calculate the error in the mean. In both panels, the dash-dotted
  line is the correlation function for the central model's
  cosmological and HOD parameters but the $\Gamma=0.12$ initial power
  spectrum. }
\end{figure}

We adopt the parameter combination $(\om, \s8) = (0.3,0.8)$ for our
central model. To populate the halos in this model, we choose
observationally motivated HOD parameters similar to those derived for
the SDSS $M_r<-20+5 \log h$ galaxy sample by \cite{zehavi04b}. The resulting
correlation function is shown by the solid line in Figure
\ref{real_space_xi}. For other $\s8$ values, we choose $M_1$,
$\alpha$, and $\mmin$ so that we closely match \xg\ of the central
model, while maintaining a fixed galaxy space density. We carry out
the HOD parameter fits using the analytic model of \xg\ described by
\cite{tinker04}, which refines the model described by
\cite{zheng04a}. The cosmological and HOD parameters of our
simulations are listed in Table 1.

We assume that satellite galaxies trace the dark matter distribution
within halos; a test in \S 3 below shows that our results are
insensitive to this assumption (see Figure \ref{hod_test}). Instead of
selecting random dark matter particles from the friends-of-friends
halos, we randomly place satellite galaxies in each halo following the
universal halo profile of Navarro, Frenk, \& White (1997; hereafter
NFW). This technique makes our results insensitive to numerical force
resolution or to discreteness effects on halo structure and velocity
dispersions. It also allows for easier comparison to analytic
approximations, since the N-body halo population is better controlled
and characterized. Most importantly for our purposes, it allows us to
choose halo concentrations appropriate to each combination of $\s8$
and $\om$, using the methods of \cite{bullock01} and
\cite{kuhlen04}.\footnote{We use the Bullock \etal\ (2001) method of
calculating $c_{\rm vir} = \rvir/r_s$, where $\rvir$ is the virial
radius of the halo and $r_s$ is the NFW scale radius. The virial
overdensity used by \cite{bullock01} depends on $\om$ and can be
significantly different from the 200 assumed here. To correct for
this, we calculate $c_{\rm vir}$ for a given halo mass $M_{\rm vir}$,
then calculate the corresponding $M_{200}$ (since the halo mass
depends on the defined edge of the halo) and scale the concentration
by $R_{200}/\rvir$. See Hu \& Kravtsov (2002) for details.} The simple
scaling of halo properties found by \cite{ztwb02} does not extend to
internal structure, which depends systematically on $\om$.  When
creating galaxy populations for models with different $\om$ but the
same $\s8$, we change halo concentrations appropriately but keep the
HOD parameters fixed. This procedure leads to small differences in
\xg\ from model to model, but these have negligible impact on our
redshift-space distortion results. We discuss concentration effects at
the end of \S 3.

We draw line-of-sight velocities of satellite galaxies (relative to
the halo center-of-mass) from a Gaussian distribution with dispersion

\begin{equation}
\sigma_v(M) = \alpha_v \left( \frac{GM}{2R_{200}} \right)^{1/2},
\end{equation}

\noindent where $R_{200}$ is the radius at which the mean density of
the halo is 200 times the background density. For $\av=1$, this choice
corresponds to the velocity distribution of an isothermal sphere.
Although a literal interpretation of $\av\ne 1$ is that the satellite
population is ``colder'' or ``hotter'' than the dark matter particles,
a modest departure from unity can also account for orbital anisotropy
and non-isothermality. In tests of anisotropy we find that a model
with one-dimensional velocity dispersions such that
$\sigma^2_i/\sigma^2_k = 2$ and $\sigma^2_j/\sigma^2_k = 0.5$, where
$i$, $j$, and $k$ are orthogonal directions randomly oriented with
respect to the axes of the box, produces quantitatively similar
results to a model with $\av=0.8$.

We use a similar technique for the velocities of central galaxies, but
here our standard assumption is that the velocity bias parameter
$\avc=0$. We also consider a model in which the central galaxies have
modest velocities, $\avc=0.2$, and an extreme model with $\avc=1$. We
also consider models with {\it satellite} $\av=0$ to isolate the
physical effects of the virial dispersion from those of the halo
velocities. The $\av=0$ models are also relevant to observational
analyses that employ ``FOG compression'', i.e., identification and
compression of galaxy groups in redshift space (see, e.g.,
\citealt{tegmark04a}). If this technique works perfectly, it
effectively sets $\av=0$ in all halos.

Figure \ref{real_space_xi} shows real-space galaxy correlation
functions for $\Gamma=0.2$ and $\s8=0.6, 0.7, 0.8, 0.9$, and $0.95$
(see Table 1 for exact values). Results are averaged over five
realizations, and error bars show the run-to-run dispersion divided by
$\sqrt{N-1}=2$ to yield the error in the mean. The inset box shows the
deviation of \xg\ for each model relative to that of the central
$(\om=0.3, \s8=0.8)$ model. The models with $\s8 \geq 0.7$ match the
central model to $\la$ 5\% at $r\la 20$ \hmpc. At larger
scales, finite box effects make the deviations larger than 10\%, but
these are smaller than the statistical errors. The $\s8=0.6$ model
matches the central model to 5\% or better at most $r$, but it
deviates by $\sim 15\%$ around 0.8 \hmpc. At roughly this scale there
is a transition between one-halo and two-halo galaxy pairs, and the
effects of $\s8$ on the halo mass function are difficult to overcome
with $P(N|M)$ changes.

The dot-dash curve in Figure \ref{real_space_xi} shows \xg\ for the
$\om=0.3$, $\s8=0.8$, $\Gamma=0.12$ model. With this large change in
the shape of the matter power spectrum, it is impossible to choose HOD
parameters that make the galaxy correlation function match that of the
$\Gamma=0.2$ models, or the SDSS data (\citealt{kev04}). Instead, for
this set of models we use the same HOD parameters found for the
corresponding $\s8$ value in the $\Gamma=0.2$ runs. The spread among
\xg\ for the five $\Gamma=0.12$ models is comparable to that for the
$\Gamma=0.2$ models. At $r<2$ \hmpc, however, the spread is
approximately twice as large.

\begin{figure*}
\centerline{\psfig{figure=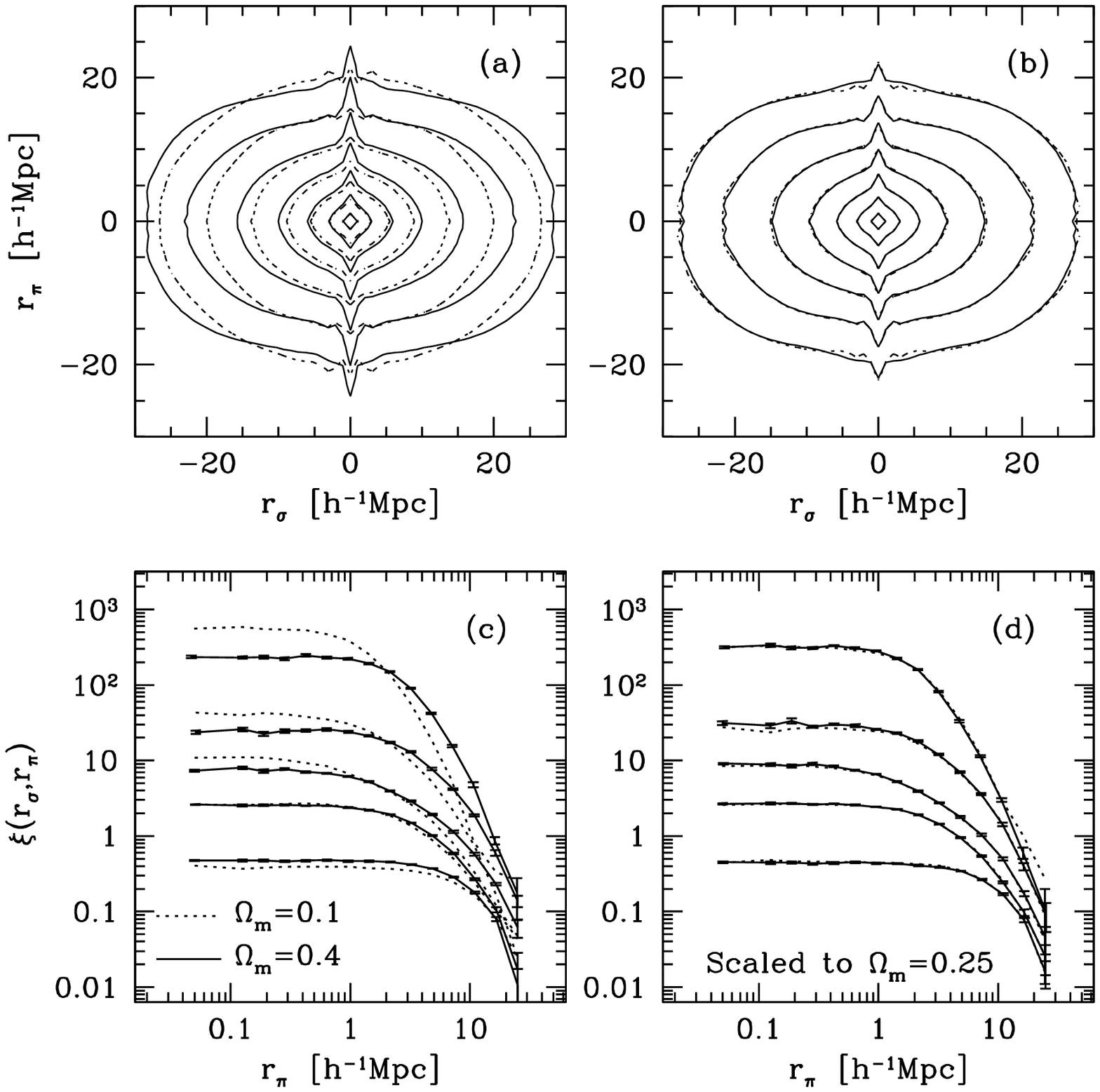,width=\two_col_fig}}
\caption{ \label{omega_scaling} Test of the $\om$-scaling
  procedure. For this test, we use simulations with $200^3$ particles
  in a $200$ \hmpc\ box and HOD parameters $\alpha=1$ and $\mmin$ and
  $M_1$ corresponding to 30 and 600 particles, respectively. (a) The
  correlation function in redshift space, \xx, for $\om=0.1$ (dotted)
  and $\om=0.4$ (solid). The contours represent lines of constant
  correlation separated by factors of two, with the outermost contour
  representing $2^{-4}$. (b) \xx\ for the same models, but now the
  galaxy velocities have been scaled to $\om = 0.25$. (c) Same models
  as (a), but now the different lines represent slices in the
  \spplane\ plane at different \rsig\ : 0.12, 0.3, 1, 3, and 11 \hmpc\
  from top to bottom. (d) The results from the two simulation sets,
  scaled to $\om=0.25$, are plotted for the same slices in the
  \spplane\ plane.  Results in all panels are averaged over three
  projections of five realizations, and error bars in (c) and (d) show
  the run-to-run dispersion divided by $\sqrt{N-1}=\sqrt{14}$ to
  calculate the error in the mean. Errors are only plotted for the
  solid lines to avoid crowding. }
\end{figure*}

\subsection{Velocity Scaling}

Figure \ref{omega_scaling} tests the efficacy of the mass/velocity
scaling technique described in \S 2.1. For this test, we ran two new
sets of \gad\ runs, each set comprised of five simulations with
$200^3$ particles in a $200$ \hmpc\ box. One set has
$(\om,\s8)=(0.1,0.8)$ at $z=0$, the other has $(\om,\s8)=(0.4,0.8)$ at
$z=0$. In both cases we chose HOD parameters $\mmin$ and $M_1$
corresponding to 30 and 600 particles, respectively, with $\alpha=1$.

Panel (a) in Figure \ref{omega_scaling} shows contours of the redshift
space correlation function, \xx, where \rsig\ represents the projected
separation between two galaxies and \rpi\ the line-of-sight
separation.  This way of representing the data is widely used in
observational studies, such as Peacock \etal\ (2001) and Hawkins
\etal\ (2003). We use the distant observer approximation, so \rpi\
simply becomes the redshift distance between galaxy pairs along one
dimension of the box, accounting for the periodic boundary
condition. Here correlation functions are averaged over three
projections of five realizations for a total of fifteen
measurements. The higher density, $\om=0.4$ model shows stronger
compression of contours at large scales because of larger coherent
flows, and it shows stronger FOG distortions at small scales because
of larger dispersions between and within halos.

Panel (c) presents the same data in a different fashion. Each line in
the panel represents the value of \xx\ as a function of \rpi\ at a
given \rsig, a slice in the \spplane\ plane. At $r_\sigma=0.12$ \hmpc,
the $\om=0.4$ model starts at a lower value of \xx\ but remains
horizontal for a longer range of \rpi. The extended horizontal plateau
reflects the longer FOGs in the higher density model, and since the
pairs at small \rsig\ are spread over a larger range of \rpi, the
amplitude near $r_\pi=0$ is necessarily depressed. We will use the
turnover of $\xi(r_\pi)$ at small \rsig\ as a quantitative measure of
small-scale distortions in \S 5. At the bottom of panel (c), where the
lines represent $r_\sigma = 11$ \hmpc, the $\om=0.4$ line is above
the $\om=0.1$ line because of the large amplification of clustering in
the coherent infall regime.

In the right panels, (b) and (d), we have scaled the velocities of the
halo and galaxy populations of both models to $\om=0.25$ in the manner
described in \S 2.1, keeping HOD parameters fixed in {\it particle
number} (and thus scaled in mass proportional to $\om$). In both
manners of representing the data, the correlation functions are nearly
indistinguishable. In other words, we can scale an $\om=0.1$ model to
$\om=0.25$ and $\om=0.4$ model to $\om=0.25$ and get the same
result. Figure \ref{omega_scaling} demonstrates that our velocity
scaling technique can be applied to our simulations without accruing
systematic errors at either large or small scales.


\begin{figure*}
\centerline{\psfig{figure=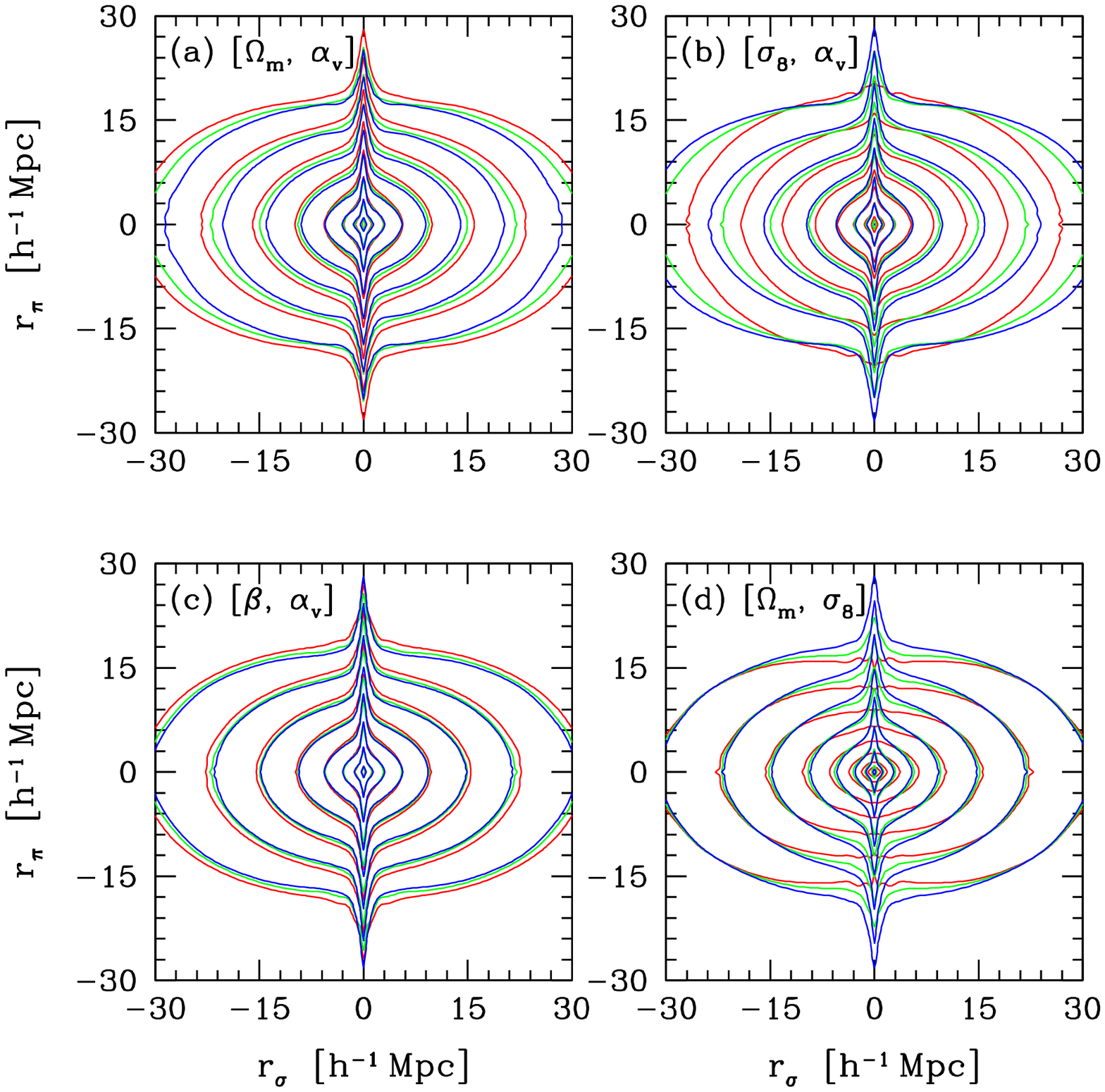,width=\two_col_fig}}
\caption{ \label{contour} Contour plots of \xx\ in the \spplane\
  plane, for the four sets of simulations listed in Table 2. The
  contours are separated by factors of two in \xx, with the outermost
  contours representing \xx $=2^{-4}$. (a) Models with $\om=0.3$,
  $\av=1$, and $\s8=0.95$ ({\it red}), $\s8=0.80$ ({\it green}),
  $\s8=0.60$ ({\it blue}). (b) Models with $\s8=0.8$, $\av=1$, and
  $\om=0.1$ ({\it red}), $\om=0.3$ ({\it green}), $\om=0.5$ ({\it
  blue}). (c) Models with $\beta\equiv\om^{0.6}/b_g=0.46$, $\av=1$, and
  $\s8=0.95$ ({\it red}), $\s8=0.80$ ({\it green}), $\s8=0.60$ ({\it
  blue}). (d) Models with $\om=0.3$, $\s8=0.8$, and $\av=0$ ({\it
  red}), $\av=0.8$ ({\it green}), $\av=1.2$ ({\it blue}). }
\end{figure*}

\begin{figure*}
\centerline{\psfig{figure=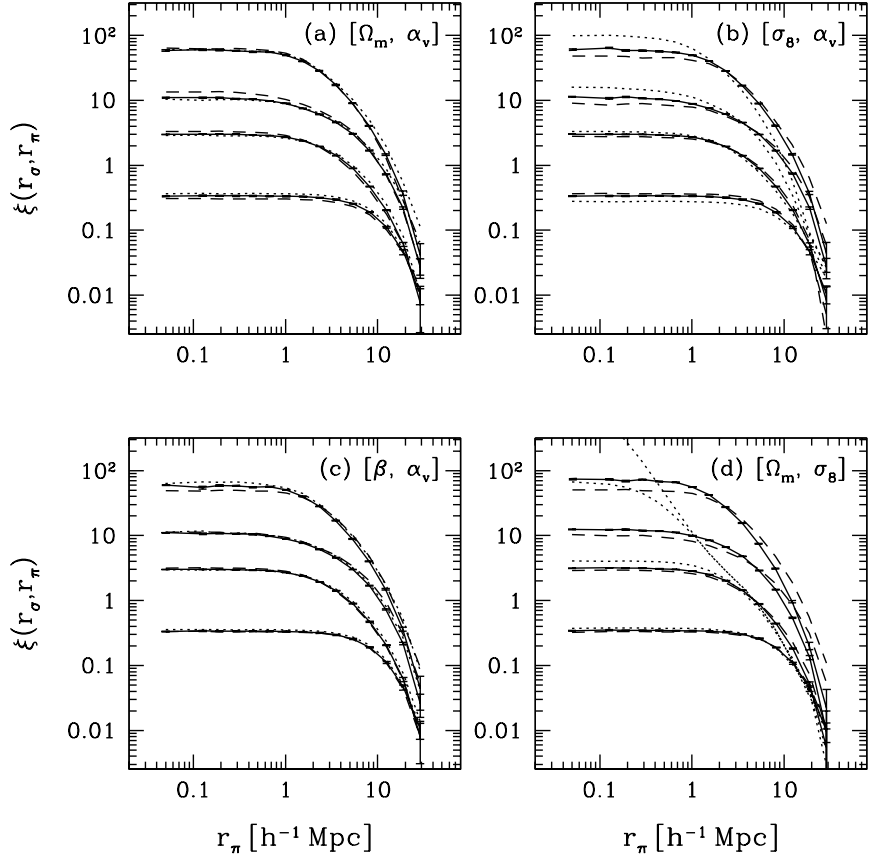,width=\two_col_fig}}
\caption{ \label{sigma_pi} Redshift-space correlation functions for
  the same models as in Figure \ref{contour}, now plotted in the slice
  format, which allows easier identification of the small-scale
  distortion. In each panel, curves represent $r_\sigma=0.12$, 0.46,
  2.54, and 14.2 \hmpc\ from top to bottom. (a) Model with $\om=0.3$,
  $\av=1$, and $\s8=0.95$ ({\it dotted}), $\s8=0.80$ ({\it solid}),
  $\s8=0.60$ ({\it dashed}). (b) Models with $\s8=0.8$, $\av=1$, and
  $\om=0.1$ ({\it dotted}), $\om=0.3$ ({\it solid}), $\om=0.5$ ({\it
  dashed}). (c) Models with $\beta\equiv\om^{0.6}/b_g=0.46$, $\av=1$,
  and $\s8=0.95$ ({\it dotted}), $\s8=0.80$ ({\it solid}), and
  $\s8=0.60$ ({\it dashed}). (d) Models with $\om=0.3$, $\s8=0.8$, and
  $\av=0$ ({\it dotted}), $\av=0.8$ ({\it solid}), $\av=1.2$ ({\it
  dashed}). Errors are the run-to-run dispersion divided by
  $\sqrt{14}$ to calculate the error in the mean. Error bars are only
  plotted for the solid lines to avoid crowding. }
\end{figure*}

\begin{figure*}
\centerline{\psfig{figure=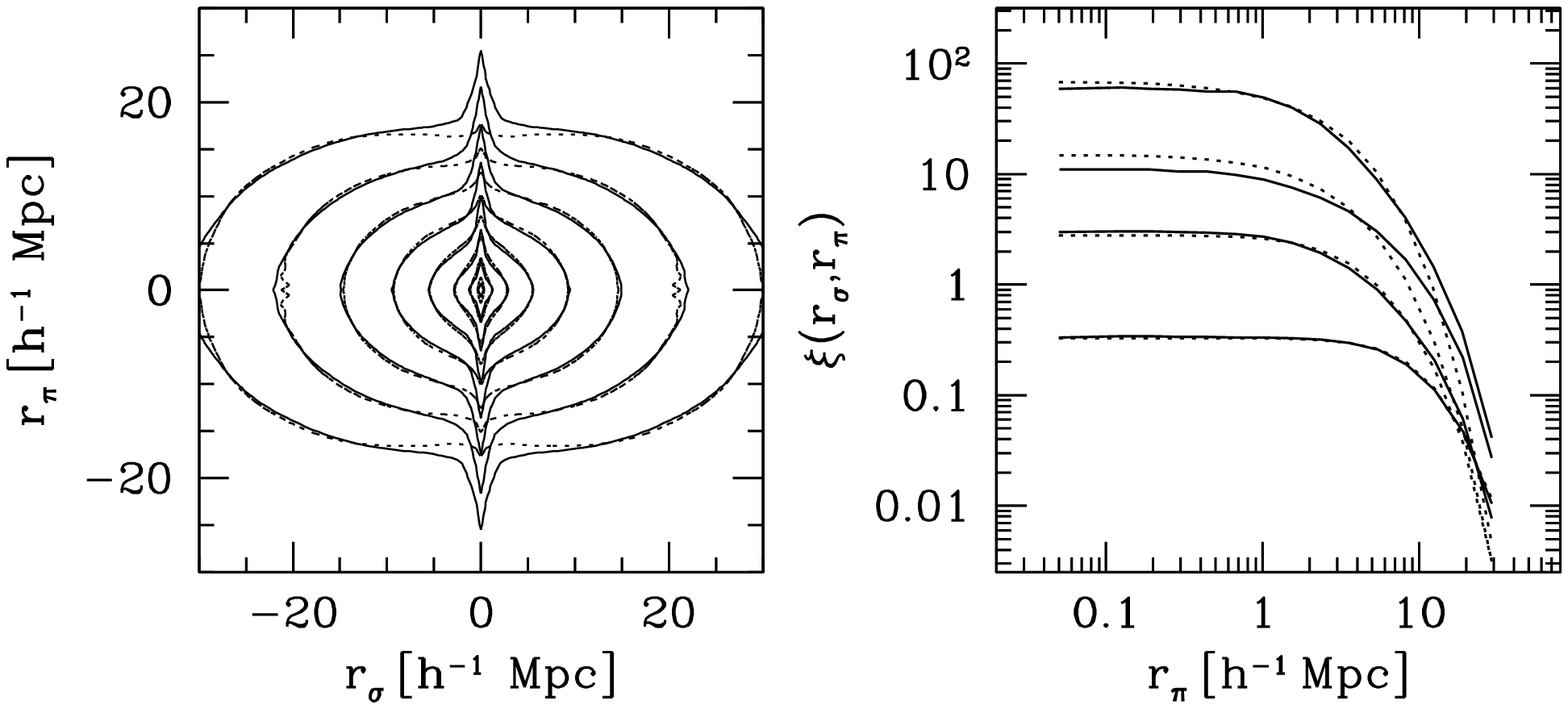,width=\two_col_fig}}
\caption{ \label{kaiser_compare} Numerical results for the central
  model $(\om=0.3,\s8=0.8,\av=1)$ compared to the best fit
  linear-exponential model. The solid lines are the numerical results
  and the dotted lines are the model. To fit the data, the exact value
  of $\beta$ was used, and the best fit dispersion $\sigma_8=418$
  \kms\ was found by $\chi^2$ minimization. Contour levels and \rsig\
  values are the same as those used in Figures \ref{contour} and
  \ref{sigma_pi}. }
\end{figure*}

\section{Overview of \xx}

Figure \ref{contour} encapsulates the dependence of the redshift-space
correlation function, \xx, on position in the $(\om,\s8,\av)$
parameter space. Each panel shows contours of \xx, separated by factors
of two, for a sequence of models in which two parameters or parameter
combinations are held fixed and one is allowed to vary. Recall that
these variations in cosmological parameters or velocity bias are
carried out at fixed (or nearly fixed) {\it real-space} galaxy
clustering, as shown in Figure \ref{real_space_xi}. The green contours
in each panel of Figure \ref{contour} show the central model with
$\om=0.3$, $\s8=0.8$, $\av=1.0$, $\avc=0$, and all models have
$\Gamma=0.2$.

In panel (a), blue and red contours show models with $\s8=0.6$ and
0.95, respectively, still with $\om=0.3$ and $\av=1$. As $\s8$
increases, \xx\ contours become more flattened because the amplitude
of coherent flows increases with larger dark matter fluctuations. In
terms of equation (\ref{e.kaiser}), higher $\s8$ means a lower galaxy
bias factor for fixed galaxy clustering amplitude, and thus a higher
value of $\beta=\om^{0.6}/\bg$. In the FOG regime at small \rsig,
contours of the three models are nearly degenerate at $r_\pi \le 10$
\hmpc. At these scales, most galaxy pairs are common members of
intermediate mass halos, and the FOG distortion depends on the masses
of those halos. The halo mass function is only weakly dependent on
$\s8$ at these intermediate masses, so the contours converge. However,
a high-$\s8$ model has more high mass halos with large virial velocity
dispersions, so at large \rpi\ the contours extend further for higher
$\s8$.

Figure \ref{contour}b shows a model sequence in which $\s8=0.8$,
$\av=1$, and $\om=0.1$ (red), 0.3 (green), and 0.5 (blue). The
flattening of contours at large \rsig\ and elongation at small \rsig\
both increase with $\om$, since a higher density universe has larger
amplitude coherent flows and more massive halos. While the large scale
distortions have a similar qualitative dependence on $\s8$ and $\om$,
the FOG distortions show an important difference. Changing $\om$
shifts the halo mass function coherently at all masses, but changing
$\s8$ shifts the high and low ends of the mass function in opposite
directions, with little change at intermediate masses. As a result,
the FOG contours converge for the varying $\s8$ sequence in panel (a)
but not for the varying $\om$ sequence in panel (b).

In panel (c), we again vary $\s8$ from 0.6 to 0.8 to 0.95, but for
each value of $\s8$ we choose the value of $\om$ that keeps the
combination $\beta=\om^{0.6}/\bg$ constant. Note that $\bg$ is
approximately proportional to $\s8^{-1}$, so this sequence has
approximately constant $\s8\om^{0.6}$, but not exactly (see \S 4.4 for
further discussion). Here the contours overlap almost perfectly on
large and intermediate scales, and they are similar even in the FOG
regime. While linear theory does not predict the form of \xx\
accurately even on the largest scale shown (see Figure
\ref{kaiser_compare} below), it correctly predicts that the class of
models with constant $\beta$ is nearly degenerate with respect to
redshift-space distortions. The differences in the FOG regime, though
difficult to see on this Figure, will nonetheless prove sufficient to
distinguish models with the same $\beta$ but different $\s8$.

In panel (d) we explore the effect of velocity bias. This sequence
uses the central values of $\om=0.3$ and $\s8=0.8$, (and thus has
constant $\beta$), with $\av$ equal to 0, 0.8, and 1.2. For
clarity, we omit the $\av=1$ model from the plot. The $\av=0$ model,
which would represent measurements from a data set with perfect ``FOG
compression,'' has elliptical contours at all scales, with no trace of
the elongation at small \rsig. Since velocity bias is applied only
within halos, these contours show that FOG distortions in \xx\ arise
entirely from halo internal velocity dispersions. At larger scales,
the $\av=0$ model begins to coincide with the others when $r_\sigma
\ga r_\pi$. The models with $\av=0.8$ and 1.2 diverge at
approximately the same location, with higher $\av$ resulting in a
stronger FOG effect. The small scale dispersion affects any global
measure of the shape of \xx\ contours, such as quadrupole-to-monopole
ratios, but it has only a small effect at large \rsig\ {\it and}
\rpi. We have also created two models, not shown in this figure, with
no satellite velocity bias but with $\avc=0.2$ and 1. These models
will be discussed in subsequent sections.

For the remainder of the paper, we will refer to these four model
sequences by writing the parameters that are held constant in square
brackets. Panel (a) plots the \ma\ sequence, panel (b) plots the \mb\
sequence, panel (c) plots the \mc\ sequence, and panel (d) plots the
\md\ sequence. The values of $\om$, $\s8$, $\beta$, $\av$, and $\avc$
for these four model sequences are listed in Table 2.

Figure \ref{sigma_pi} plots the same results as Figure \ref{contour},
but now in the form of Figure \ref{omega_scaling}, showing slices at
fixed values of \rsig. For each model, the top two curves trace out
the FOG distortions at $r_\sigma \approx 0.12$ \hmpc\ and
$r_\sigma=0.46$ \hmpc, allowing discrimination of models in the FOG
regime that is difficult from the contour plots alone.

In panel (a), changes in $\s8$ at fixed \ma\ have only a small effect
on the FOG distortions at $r_\sigma = 0.12$ \hmpc, though even these
changes are significant relative to our statistical error bars, which
are comparable to the line width. At $r_\sigma = 14$ \hmpc, the
high-$\s8$ model has higher \xx\ at all \rpi, but the large scale
distortions are more difficult to discriminate in this representation
compared to the contour plot (Fig. \ref{contour}a).

\begin{figure*}
\centerline{\psfig{figure=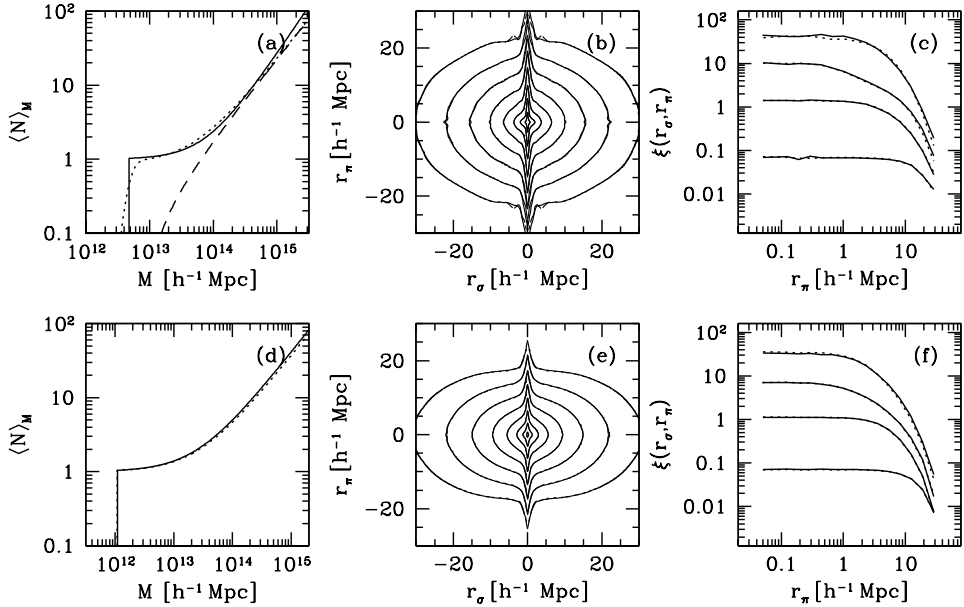,width=\two_col_fig}}
\caption{ \label{hod_test} The influence of HOD parameterization or
  halo concentration on predicted redshift-space distortions, when HOD
  parameters are chosen to yield the same real-space galaxy
  correlation function. All models assume $\om=0.3$, $\s8=0.8$,
  $\av=1$. The top row compares results from our standard
  three-parameter HOD to the five-parameter HOD of Zheng et al.\
  (2004). (a) Mean occupation functions $\navg$ for the
  three-parameter HOD (solid line) and the five-parameter HOD (dotted
  line). The dashed curve is the satellite contribution to $\navg$ for
  the five-parameter HOD. (b) \xx\ contour plots for the
  three-parameter HOD (solid line) and the five-parameter HOD (dotted
  line, virtually indistinguishable). (c) Slices in \spplane\ for the
  three-parameter HOD (solid lines) and five-parameter HOD (dotted
  lines). (d) --- (f): Similar to (a) --- (c), but dotted curves
  represent a model in which concentrations have been reduced by 30\%
  and the HOD parameters (in the three-parameter model) have been
  adjusted to maintain the small-scale correlation function. }
\end{figure*}

In the remaining panels, parameter changes have a marked effect on the
FOG distortions at small \rsig. In particular, the models with
constant \mc, which have nearly identical large scale distortions,
show a $\sim 40\%$ change in \xx\ at small $(r_\sigma,r_\pi)$ as
$\s8$ rises from 0.6 to 0.95 (Fig. \ref{sigma_pi}c). While the
separation of lines is not dramatic on a plot spanning five decades on
the $y$-axis, differences of tens of percent should be easily
measurable at these scales in the samples the size of the 2dFGRS and
SDSS. Changing $\av$ from 0.8 to 1.2 has an effect of similar
magnitude, though it differs in detailed form (Fig. \ref{sigma_pi}d).

Figure \ref{kaiser_compare} compares our numerical results for the
central model $(\om=0.3, \s8=0.8, \av=1)$ to the analytic,
linear-exponential model of equation (\ref{e.lin_exp}). We fix $\beta$
to the true value of 0.46 and vary $\sigma_v$ to minimize $\chi^2$ for
all data at separations larger than 10 \hmpc\ (we get similar
$\sigma_v$ if we use data at all separations). The linear-exponential
model describes the large scale distortions fairly well, though even
here there are systematic differences between the numerical \xx\
contours and the model fit. The model does a poor job of replicating
the FOG distortions at large \rpi, a failure that is evident in both
the contour plots and the line plots. These deficiencies of the
linear-exponential model can also be seen in its application to the
2dFGRS data by Peacock \etal\ (2001, see their Figure 2). There, the
measured distortions at small \rsig\ clearly extend past the model
predictions, even though the FOG effect has been smoothed relative to
our plots here by the larger bin size. We can force the
linear-exponential model to better match the FOG distortions by
adopting a higher $\sigma_v$, but the fit at large scales is then
severely degraded.

When analyzing observational data, we must infer the galaxy HOD by
fitting parameterized models to the measured real-space clustering
(e.g., the projected correlation function). We anticipate that
redshift-space distortions will be insensitive to the adopted HOD
parametrization so long as the model reproduces the observed real
space correlation function. Figure \ref{hod_test} demonstrates the
validity of this conjecture. We first populate the halos of our
$\s8=0.8$, $\om=0.3$ N-body simulations using a five-parameter HOD
model fit to results of a hydrodynamic simulation (\citealt{zheng04}),
in which the galaxy space density is $2.5\times 10^{-3} (h^{-1}{\rm
Mpc})^{-3}$. This parameterization incorporates adjustable smooth
cutoffs in the central and satellite galaxy mean occupation functions,
and it can achieve an essentially perfect fit to the predictions of
semi-analytic and numerical models of galaxy formation
(\citealt{zheng04}). We then fit parameters of our restricted,
three-parameter HOD model to reproduce the correlation function of the
five-parameter model as closely as possible, obtaining agreement
similar to that in Figure \ref{real_space_xi}. Figure \ref{hod_test}a
shows the original and fitted mean occupation functions, and Figures
\ref{hod_test}b and \ref{hod_test}c show \xx\ for the two models, in
the format of Figures \ref{contour} and \ref{sigma_pi},
respectively. While the sharp cutoff model cannot represent the
$\navg$ of the input model exactly, it predicts essentially
indistinguishable redshift-space distortions. The large scale
distortions for both models are weaker than those in Figures
\ref{contour} and \ref{sigma_pi} because our HOD parameters are
matched to a strongly clustered galaxy sample with higher $b_g$ and
consequently lower $\beta$.

As discussed in \S 2.2, our HOD models assume that satellite galaxies
in halos have the same radial profile as the dark matter. If we change
this assumption when fitting the observed correlation function, or if
we make this assumption but it does not hold in the real universe,
then we will derive slightly different HOD parameters, which in turn
will change the redshift-space distortions. We test our sensitivity to
the radial profile assumption by creating a model that matches \xg\ of
our standard central model but uses satellite profile concentrations
$30\%$ lower than those of the dark matter halos themselves. Figure
\ref{hod_test}d shows the mean occupation functions of the two
models. The low concentration model has a lower $M_1$ to create more
close one-halo pairs, and a lower $\alpha$ to prevent overpopulation
of massive halos. Figures \ref{hod_test}e and \ref{hod_test}f show the
redshift-space distortions of the two models. The large scale
distortions of the two models are the same, apparent from both the
contour plots and the line plots. The low concentration model has
slightly weaker fingers-of-god because it has fewer galaxies in
massive halos, but this difference is barely distinguishable in Figure
\ref{hod_test}f, and the difference in the quantitative measures of
small scale distortion measures introduced in \S 5 is within our
statistical errors. We conclude that departures from the standard
radial profile by $\le 30\%$ do not alter our results. Still larger
changes might have noticeable effect, since the inferred HODs would
predict different non-linear velocity fields, but substantial
departures from theoretically predicted dark matter profiles can be
detected observationally by measuring satellite galaxy profiles in
groups and clusters.


\section{Measures of Large-Scale Distortion and the Value of $\beta$}

The blueprint for cosmological parameter estimation begins at large
scales. At these scales, anisotropies are governed by the value of
$\beta=\om^{0.6}/\bg$ (see Figure \ref{contour}). The effects of
velocity bias are limited and, we will show, straightforward to
remove. Values of $\bg$ for our five values of $\s8$ are listed in
Table 1. We define galaxy bias factors by the ratio of the non-linear,
real-space galaxy and matter correlation functions in the range $4 \le
r \le 12$ \hmpc, $\bg^2\equiv \xi_g/\xi_m$, a choice that we discuss
further in \S 4.4 below.  Changing the range to $10\le r \le 25$
\hmpc\ changes the values by $\la 1\%$. In characterizing
distortions of the power spectrum or correlation function, we follow
the track of Kaiser (1987), Hamilton (1992), and Cole \etal\ (1994),
using either the ratio of the angle-averaged redshift-space quantity
to the real-space quantity, or the ratio of the quadrupole moment to
the monopole in redshift space. The two methods applied to two
statistics provide four measures of large scale distortions,
illustrated by Figures \ref{pk_zr}--\ref{xi_qp} below.

\subsection{The Power Spectrum}

The angular dependence of the redshift-space galaxy power spectrum can
be characterized as a sum of Legendre polynomials, denoted here as
$L_l(\mu)$,

\begin{equation}
\label{e.PZ_sum}
P_Z(k,\mu) = \sum_{l=0}^\infty P_l(k) L_l(\mu).
\end{equation}

\noindent This equation can be inverted to determine each individual
multipole by

\begin{equation}
\label{e.P_l}
P_l(k) = \frac{2l+1}{2} \int_{-1}^{+1} P_Z(k,\mu) L_l(\mu) d\mu.
\end{equation}

\noindent Statistical symmetry of positive and negative peculiar
velocities guarantees that odd multipoles vanish on average. In linear
perturbation theory, only the $l=0,2$, and $4$ moments are
non-zero. Equations (\ref{e.kaiser}) and (\ref{e.P_l}) yield

\begin{equation}
\label{e.P0}
P_0(k) = \left(1+\frac{2}{3}\beta + \frac{1}{5}\beta^2\right) P_R(k),
\end{equation}

\begin{equation}
\label{e.P2}
P_2(k) = \left(\frac{4}{3}\beta + \frac{4}{7}\beta^2\right) P_R(k),
\end{equation}

\noindent for the monopole and the quadrupole, where $P_R(k)$ is the
real-space power spectrum. In linear theory, the angle-averaged
redshift-space power spectrum $P_0(k)$ is amplified over the
real-space power spectrum by a constant factor, and the enhancement of
fluctuations along the line of sight produces a positive quadrupole
$P_2(k)$ with the same shape as $P_R(k)$. The ratio of the monopole to
the real-space power spectrum, $\pkzr$, or the quadrupole-to-monopole
ratio, $\pkqp$, are scale-independent functions of $\beta$:

\begin{equation}
\label{e.pkzr}
\pkzr(k) \equiv \frac{P_0(k)}{P_R(k)} = 1+\frac{2}{3}\beta + \frac{1}{5}\beta^2\,,
\end{equation}

\begin{equation}
\label{e.pkqp}
\pkqp(k) \equiv \frac{P_2(k)}{P_0(k)} = \frac{\frac{4}{3}\beta + \frac{4}{7}\beta^2}
{1+\frac{2}{3}\beta + \frac{1}{5}\beta^2}\,\,.
\end{equation}

\noindent However, non-linear effects, especially the velocity
dispersions in collapsed or collapsing structures, suppress $P_0(k)$
at smallest scales and cause the quadrupole to actually reverse sign in
the non-linear regime. In practice, the ratios $\pkzr$ and $\pkqp$ are
monotonically decreasing functions of $k$, and equations
(\ref{e.pkzr}) and (\ref{e.pkqp}) do not provide accurate estimates of
$\beta$ at scales accessible to high-precision measurements. The use
of the linear-exponential model (eq. \ref{e.lin_exp}) in place of pure
linear theory (eq.\ \ref{e.kaiser}) can greatly improve the accuracy of
$\beta$ estimates, but it still does not remove biases entirely
(\citealt{cfw95,hc99}).

\begin{figure*}
\centerline{\psfig{figure=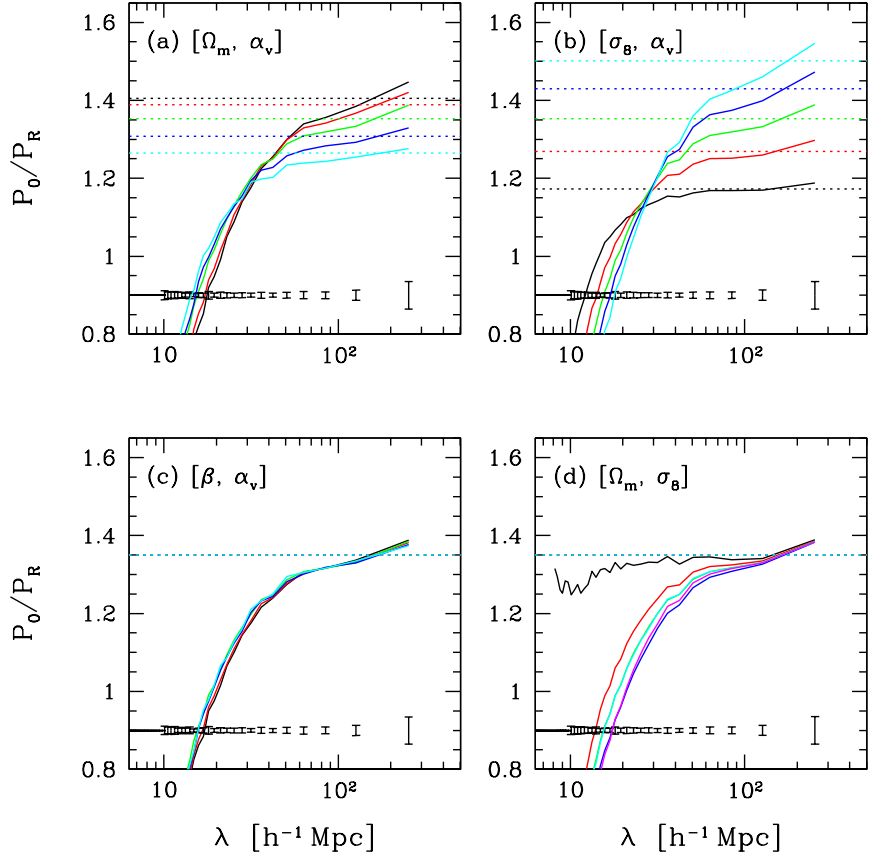,width=\two_col_fig}}
\caption{ \label{pk_zr} Monopole-to-real space ratio of the power
  spectrum, $P_0/P_R$, as a function of wavelength, up to the
  fundamental mode of the 253 \hmpc\ box. The dotted lines represent
  the linear theory prediction for each model for this measure. The
  error bars plotted at the bottom of each panel are errors in the
  mean for the central model, which are generally representative of
  the error bars for the rest of the models in each set. Models run
  from $\s8=0.95$ ({\it black}) to $\s8=0.6$ ({\it cyan}) in panel
  (a), from $\om=0.1$ ({\it black}) to $\om=0.5$ ({\it cyan}) in panel
  (b), and from $\s8=0.95$ ({\it black}) to $\s8=0.6$ ({\it cyan}) in
  panel (c), with the order of the colors being black, red, green,
  blue, cyan. In panel (d), models are $\av=0$ ({\it black}),
  $\av=0.8$ ({\it red}), $\av=1.0$ ({\it green}), $\av=1.2$ ({\it
  blue}), and $\avc=0.2$ ({\it cyan}, barely separable from green),
  and $\avc=1$ ({\it magenta}). }
\end{figure*}

\begin{figure*}
\centerline{\psfig{figure=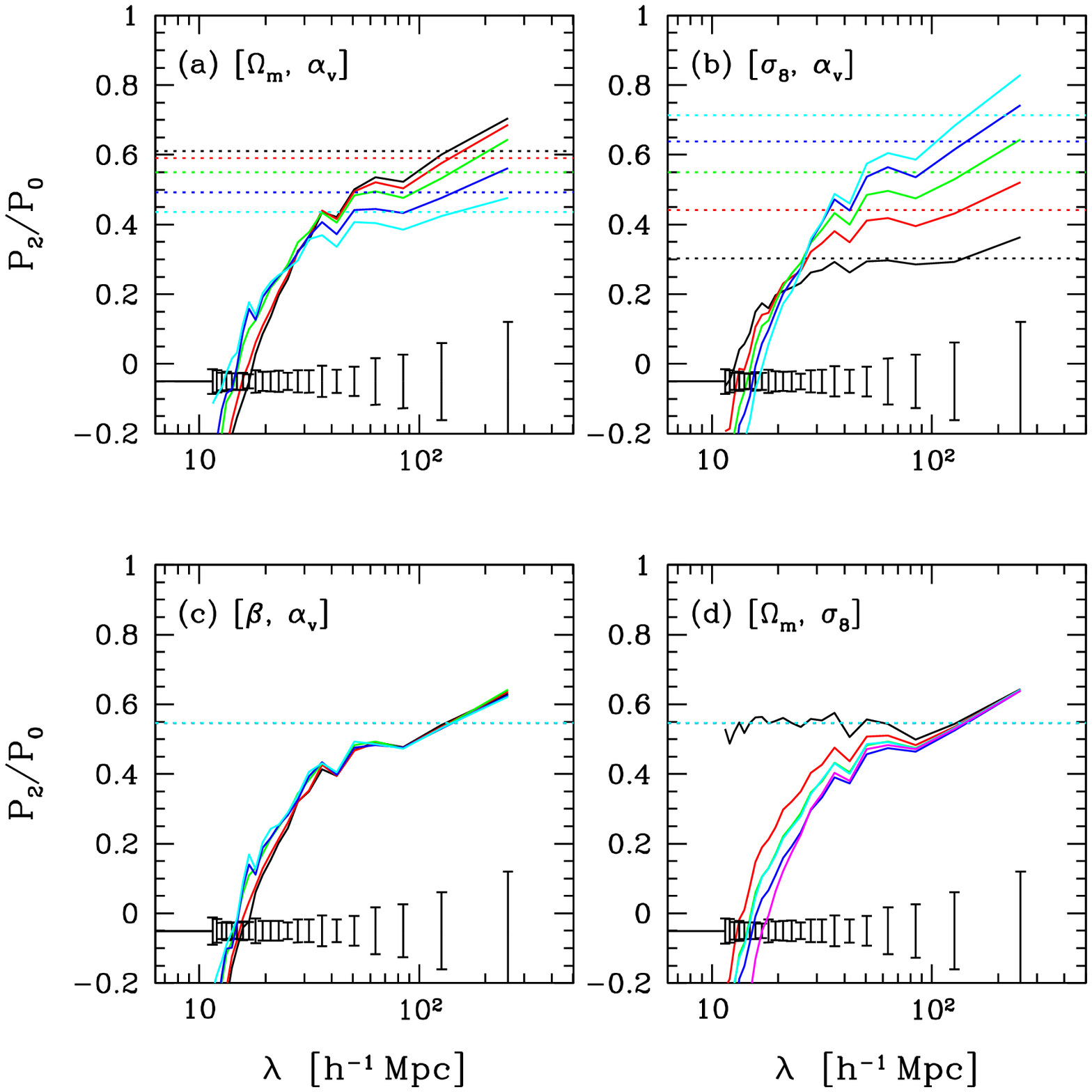,width=\two_col_fig}}
\caption{ \label{pk_qp} Quadrupole-to-monopole ratio of the redshift
  space power spectrum, $P_2/P_0$, as a function of wavelength, up to
  the fundamental mode of the 253 \hmpc\ box. The dotted lines and
  error bars are as in Figure \ref{pk_zr}. Models run from $\s8=0.95$
  ({\it black}) to $\s8=0.6$ ({\it cyan}) in panel (a), from $\om=0.1$
  ({\it black}) to $\om=0.5$ ({\it cyan}) in panel (b), from
  $\s8=0.95$ ({\it black}) to $\s8=0.6$ ({\it cyan}) in panel (c),
  with the order of the colors being black, red, green, blue, cyan. In
  panel (d), models are $\av=0$ ({\it black}), $\av=0.8$ ({\it red}),
  $\av=1.0$ ({\it green}), $\av=1.2$ ({\it blue}), and $\avc=0.2$
  ({\it cyan}, barely separable from green), and $\avc=1$ ({\it
  magenta}). }
\end{figure*}

To calculate the redshift-space galaxy power spectra for our
simulations, we use the same technique as Berlind, Narayanan, \&
Weinberg (2001). In the distant observer approximation, we take an
axis of the box as the line of sight, wrap particles around the
periodic boundary if their peculiar velocities shift them outside the
box, and calculate $\pzkmu$ by Fast Fourier Transform. We use a
$200^3$ density mesh and treat each axis as an independent line of
sight. The multipole moments are calculated by fitting the first three
even terms in equation (\ref{e.P_l}).  We compute the average from 15
measurements (three projections of five simulations) and the errors by
dividing the run-to-run dispersions by $\sqrt{14}$. Figures
\ref{pk_zr} and \ref{pk_qp} show the results of this analysis for
$\pkzr$ and $\pkqp$, respectively, as functions of wavelength
$\lambda=2\pi/k$. Horizontal dotted lines represent the values of
$\pkzr$ and $\pkqp$ predicted by linear theory (eqs. \ref{e.pkzr} and
\ref{e.pkqp}).

Figures \ref{pk_zr}a and \ref{pk_zr}b plot $\pkzr (k)$ for varying
$\s8$ and $\om$, respectively. At large $\lambda$, $\pkzr$ increases
with increasing $\beta$. But all the curves drop rapidly at scales
$\lambda \sim 30$ \hmpc\ due to non-linearities. The difficulty in
using linear theory to extract $\beta$ is easily seen; none of the
models shows a clear asymptotic value of $\pkzr$. An estimate of the
linear theory value might be possible for the lowest value of $\om$ or
$\s8$, but as either parameter increases the slope of the curve at
large $\lambda$ becomes larger. At $\beta\ga 0.4$ the data never
converge to the large-scale horizontal asymptote predicted by linear
theory, even at the fundamental mode of the box.

For constant \mc, in panel (c), the curves are nearly identical within
the error bars, especially at large scales. Thus, even though linear
theory does not yield an accurate estimate of $\beta$, it predicts the
{\it scaling} of $\pkzr$ with cosmological parameters almost
perfectly, quantifying the visual impression of Figure
\ref{contour}c. In panel (d), the behavior of the $\av=0$ model
demonstrates that random dispersion in virialized groups plays a
dominant role on suppressing $\pkzr$. With the virial motions
eliminated, the data for this model remain nearly constant over more
than a decade in $\lambda$, with the other curves only meeting it at
$\lambda \sim 100$ \hmpc. A sufficiently effective FOG compression
technique might therefore allow useful estimation of $\beta$ from
linear theory and $\pkzr$.

The other velocity bias models begin to diverge from each other at
$\lambda \sim 70$ \hmpc, again demonstrating that cluster virial
velocities affect redshift distortions well into what is normally
considered the linear regime. If we allow central galaxies to move
with respect to the halo center-of-mass with bias $\avc=0.2$, we find
barely detectable changes (the line cannot be seen because it is
directly beneath the line for the central model). We also plot the
model with $\avc=1$, in which the central galaxy random velocities are
the same magnitude as those of dark matter particles. At small scales,
adding large central galaxy velocities has roughly the same effect as
increasing the {\it satellite} velocity bias to $\av=1.2$, but the
$\avc=1$ model converges with the central model somewhat faster.

Results for the quadrupole-to-monopole ratio are shown in Figure
\ref{pk_qp}. The model dependence of $\pkqp$ is qualitatively similar
to that of $\pkzr$, though the use of a higher order multipole leads
to substantially larger statistical error. As with $\pkzr$, the
$\pkqp$ curves only reach a large scale asymptote for the lowest
values of $\beta$. Once again, however, linear theory correctly
predicts that models with constant $\beta$ have the same large scale
distortions.  For the fixed \md\ model set in panel (d), the $\av=0$
model is consistent with linear theory at $\lambda > 20$
\hmpc. Increasing satellite velocity dispersions suppresses $\pkqp$ at
steadily larger scales. Central galaxy velocities with $\avc=0.2$
produce almost no change, while the model with $\avc=1$ shows even
stronger suppression than the satellite $\av=1.2$ model.

\subsection{The Correlation Function}

Since the power spectrum and correlation function are related by
Fourier transformation, the linear theory approximation to
$P_Z(k,\mu)$ also applies to \xx. Hamilton (1992) introduced the
multipole approximation in configuration space, devising linear
theory diagnostics of \xx\ that parallel those in equations
(\ref{e.pkzr}) and (\ref{e.pkqp}). The multipoles of the redshift
space correlation function, $\xi_l(r)$, are calculated by the same
inversion formula used in the Fourier domain,

\begin{equation}
\label{e.xi_l}
\xi_l(r) = \frac{2l+1}{2} \int_{-1}^{+1} \xi(r_\sigma,r_\pi) L_l(\mu) d\mu,
\end{equation}

\noindent where $r=\sqrt{r_\sigma^2 + r_\pi^2}$ and $\mu=r_\pi/r$. The
ratio of the monopole, $\xi_0(r)$, to the real-space correlation
function, \xg, exactly parallels equation (\ref{e.pkzr}),

\begin{equation}
\label{e.xizr}
\xizr(r) \equiv \frac{\xi_0(r)}{\xi_R(r)} = 1 + \frac{2}{3}\beta + \frac{1}{5}\beta^2.
\end{equation}

\noindent The quantity

\begin{equation}
\label{e.xiqp}
\xiQp(r) \equiv \frac{\xi_2(r)}{\xi_0(r) - \bar{\xi}_0(r)} = 
\frac{\frac{4}{3}\beta + \frac{4}{7}\beta^2}{1+\frac{2}{3}\beta + \frac{1}{5}\beta^2},
\end{equation}

\noindent has the same asymptotic value as $\pkqp$ in linear theory
(assumed for the second equality above). Here $\bar{\xi}_0(r)$ is
the spherically averaged monopole,

\begin{equation}
\label{e.xi_bar}
\bar{\xi}_0(r) = \frac{3}{r^3}\int_0^r\xi_0(s) s^2 ds.
\end{equation}

\noindent We henceforth refer to $\xiQp$ as the quadrupole of the
redshift-space correlation function. To calculate $\xi_0(r)$ and
$\xi_2(r)$, we bin galaxy pairs on a polar grid of logarithmic spacing
in $r$ and linear spacing in angle, then perform the integral
(\ref{e.xi_l}) numerically at each $r$.

\begin{figure*}
\centerline{\psfig{figure=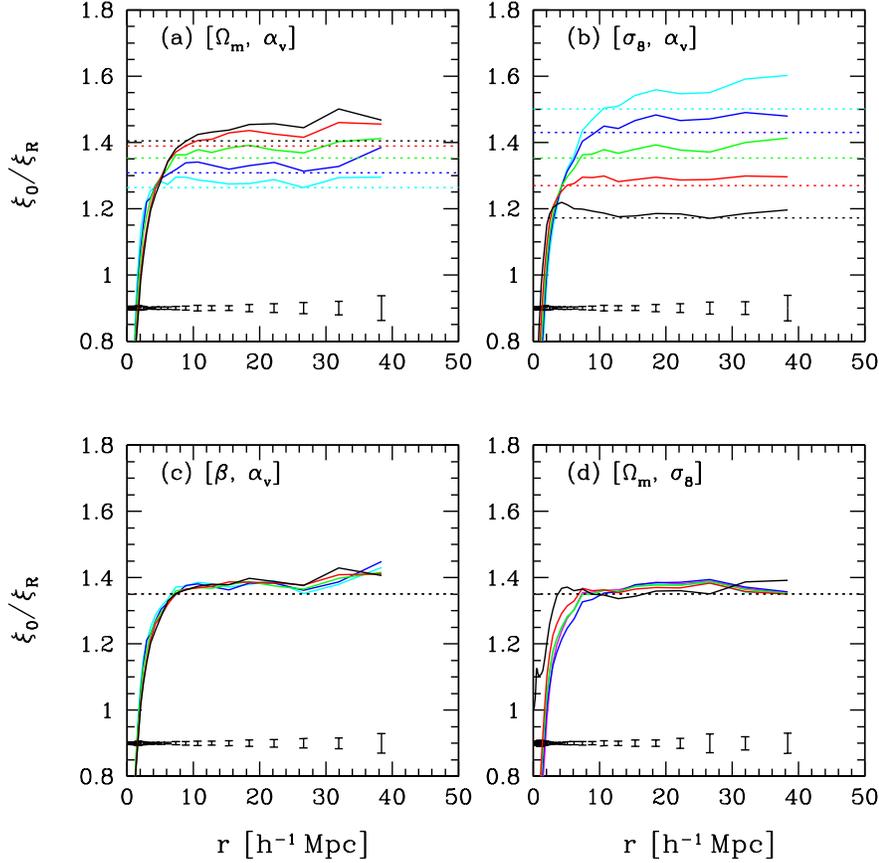,width=\two_col_fig}}
\caption{ \label{xi_zr} Ratio of the monopole of the redshift-space
  correlation function to the real-space correlation function,
  $\xizr$, as a function of separation $r$. The dotted lines and error
  bars are as in Figure \ref{pk_zr}. Models run from $\s8=0.95$ ({\it
  black}) to $\s8=0.6$ ({\it cyan}) in panel (a), from $\om=0.1$ ({\it
  black}) to $\om=0.5$ ({\it cyan}) in panel (b), from $\s8=0.95$
  ({\it black}) to $\s8=0.6$ ({\it cyan}) in panel (c), with the order
  of the colors being black, red, green, blue, cyan. In
  panel (d), models are $\av=0$ ({\it black}), $\av=0.8$ ({\it red}),
  $\av=1.0$ ({\it green}), $\av=1.2$ ({\it blue}), and $\avc=0.2$
  ({\it cyan}, barely separable from green), and $\avc=1$ ({\it
  magenta}). }
\end{figure*}

Figure \ref{xi_zr} shows the results for $\xizr$, plotted as
a linear function of $r$. In each panel, the curves reach an
asymptotic value quickly, near $r=10$ \hmpc. In most cases, the
asymptote is above the dotted line representing the linear theory
prediction. Despite this small systematic bias, which increases with
increasing $\beta$, this diagnostic does not suffer from non-linear
suppression of distortions at large scales; a fit to a constant value
is straightforward. Another notable advantage of this diagnostic is
that the effects of velocity bias (panel d) are almost negligible
beyond $r=10$ \hmpc. FOG compression ($\av=0$) removes the systematic
offset between $\xizr$ and the linear theory prediction at $r\sim
10-30$ \hmpc. This result suggests that the offset is a consequence of
FOGs transferring pairs from small separations in real space to large
separations in redshift space.

\begin{figure*}
\centerline{\psfig{figure=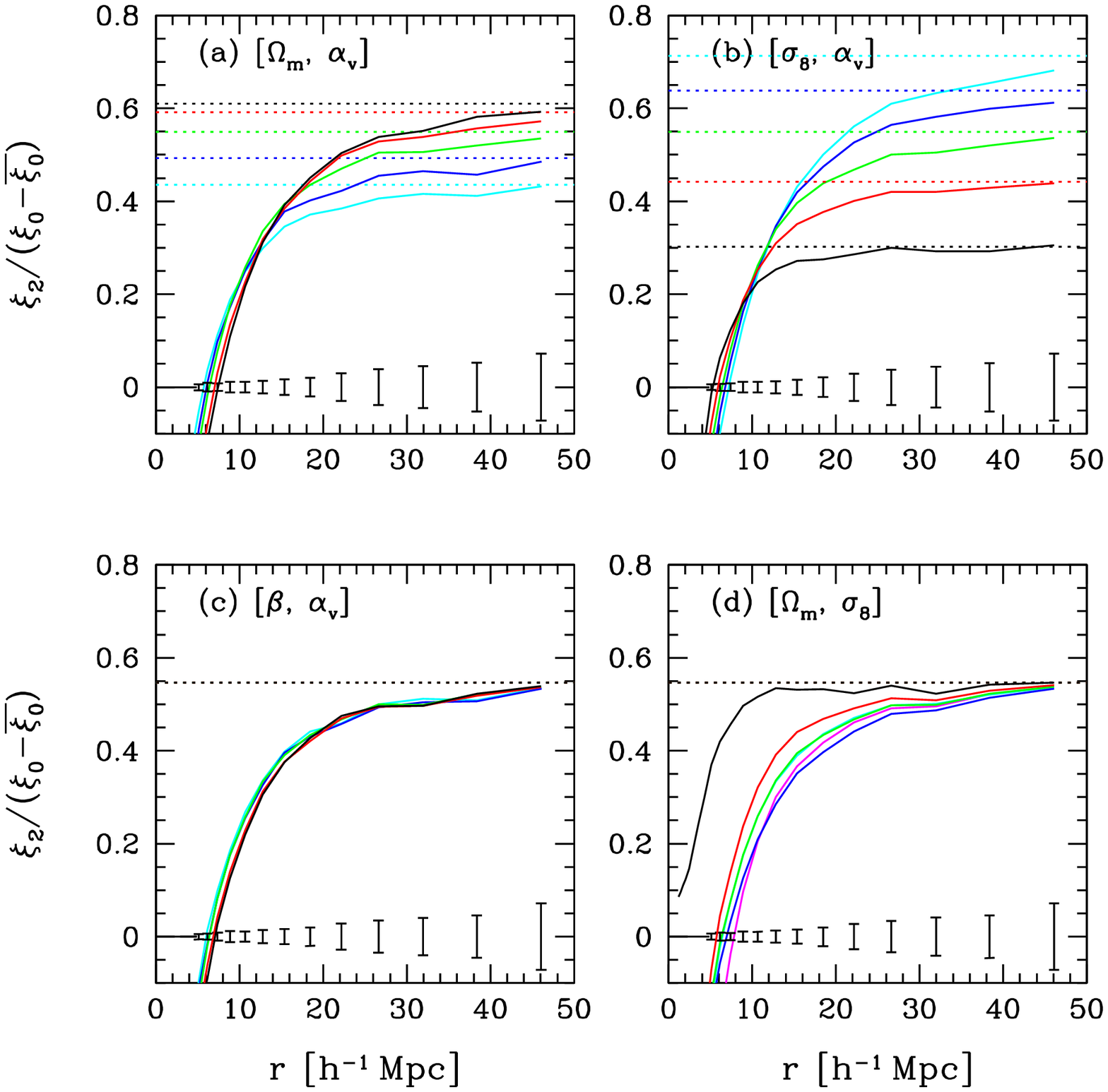,width=\two_col_fig}}
\caption{ \label{xi_qp} Quadrupole moment of the correlation function,
  $\xiQp = \xiqp$, as a function of separation.  The dotted lines and
  error bars are as in Figure \ref{pk_zr}. Models run from $\s8=0.95$
  ({\it black}) to $\s8=0.6$ ({\it cyan}) in panel (a), from $\om=0.1$
  ({\it black}) to $\om=0.5$ ({\it cyan}) in panel (b), from
  $\s8=0.95$ ({\it black}) to $\s8=0.6$ ({\it cyan}) in panel (c),
  with the order of the colors being black, red, green, blue, cyan.
  In panel (d), models are $\av=0$ ({\it black}), $\av=0.8$ ({\it
  red}), $\av=1.0$ ({\it green}), $\av=1.2$ ({\it blue}), and
  $\avc=0.2$ ({\it cyan}, barely separable from green), and $\avc=1$
  ({\it magenta}). }
\end{figure*}

Figure \ref{xi_qp} plots $\xiQp$ as a linear function of $r$. These
curves resemble those of the power spectrum measures plotted as a
function of $\log \lambda$. Models with low values of $\beta$ reach a
horizontal asymptote at large $r$, while $\xiQp$ for the high-$\beta$
models is still increasing at the largest separation. All the curves
are under the predicted linear theory values, in contrast to the
results for $\xizr$. Figure \ref{xi_qp}d shows that small scale
dispersions are the main effect suppressing $\xiQp$; with $\av=0$,
$\xiQp$ tracks the linear theory prediction down to
$r=10$\hmpc. Increasing satellite or central galaxy velocity
dispersions drives the non-linear suppression of $\xiQp$ to larger
scales.

\subsection{Estimating $\beta$}

The $\av=0$ curves in Figures \ref{pk_zr}d, \ref{pk_qp}d,
\ref{xi_zr}d, and \ref{xi_qp}d show that $\beta$ can be estimated
fairly accurately using linear theory if FOG distortions are removed
by suppressing velocity dispersions in virialized halos. However,
these curves represent a case in which FOG compression is perfect,
with halos identified in real space from the densely sampled dark
matter distribution. Any realistic scheme must operate on the sparsely
sampled galaxy distribution in redshift space, and it will suffer from
incompleteness and contamination of the halo catalog and incorrect
assignments of galaxies to halos. The impact of these imperfections on
$\beta$ estimates must be evaluated in the context of a specific group
identification scheme applied to a survey with specified depth and
geometry, and we will not consider the FOG compression approach
further in this paper. Instead, we will use our numerical results to
devise fitting procedures that estimate $\beta$ and a characteristic
non-linear scale from measurements of $\pkqp(k)$, $\pkzr(k)$,
$\xizr(r)$, and $\xiQp(r)$. In the remainder of the paper, we use the
notation $\bfit$ to represent a value of $\beta$ estimated by one of
these fitting procedures, and use $\beta$ to represent the true model
values of $\om^{0.6}/\bg$. The forms of our fitting functions are
arbitrary, motivated by efficacy rather than theoretical arguments,
but they all encode the general behavior of linear distortions at
large scales suppressed or reversed by non-linear effects at small
scales.

For the quadrupole-to-monopole ratio of the power spectrum, our
procedure is similar to that proposed by Hatton \& Cole (1999;
hereafter HC99), who suggest the fitting formula

\begin{equation}
\label{e.hc}
\pkqp(\lambda) = \pkqplin \left[ 1 - (\lambda/\lambda_0)^{-1.22}\right].
\end{equation}

\noindent Here $\pkqplin$ is the linear theory quadrupole distortion,
related to $\beta$ by equation (\ref{e.pkqp}), and $\lambda_0$ is the
non-linear scale at which the quadrupole passes through zero. We make
two changes to the HC99 procedure, which, in our experiments, improve
the accuracy and robustness of the $\beta$ estimates. First, we
calculate $\lambda_0$ by fitting a straight line to the six data
points surrounding $\pkqp=0$, instead of leaving it as a fitting
parameter in the global fit. Second, we modify equation (\ref{e.hc})
to

\begin{equation}
\label{e.fpkqp}
\pkqp(\lambda) = \pkqplin \left[1 - \left(\frac{\lambda}{\lambda_0}\right)
  ^{-1.55/(0.45+\pkqplin)}\right].
\end{equation}

\noindent We determine the fitting parameter $\pkqplin$ by minimizing
$\chi^2$ for all data points with $\lambda\ge \lambda_0$, ignoring any
covariance of errors, and we then solve for $\bfit$ using equation
(\ref{e.pkqp}). Since $\pkqplin$ varies around $\sim 0.55$, the
exponent in equation (\ref{e.fpkqp}) is similar to that in HC99's
formula, but including a dependence on $\pkqplin$ captures the
behavior seen in Figure \ref{pk_qp}, where the $\pkqp$ curves for
higher $\beta$ models flatten toward their asymptotic values at
larger scales.

\begin{figure*}
\centerline{\psfig{figure=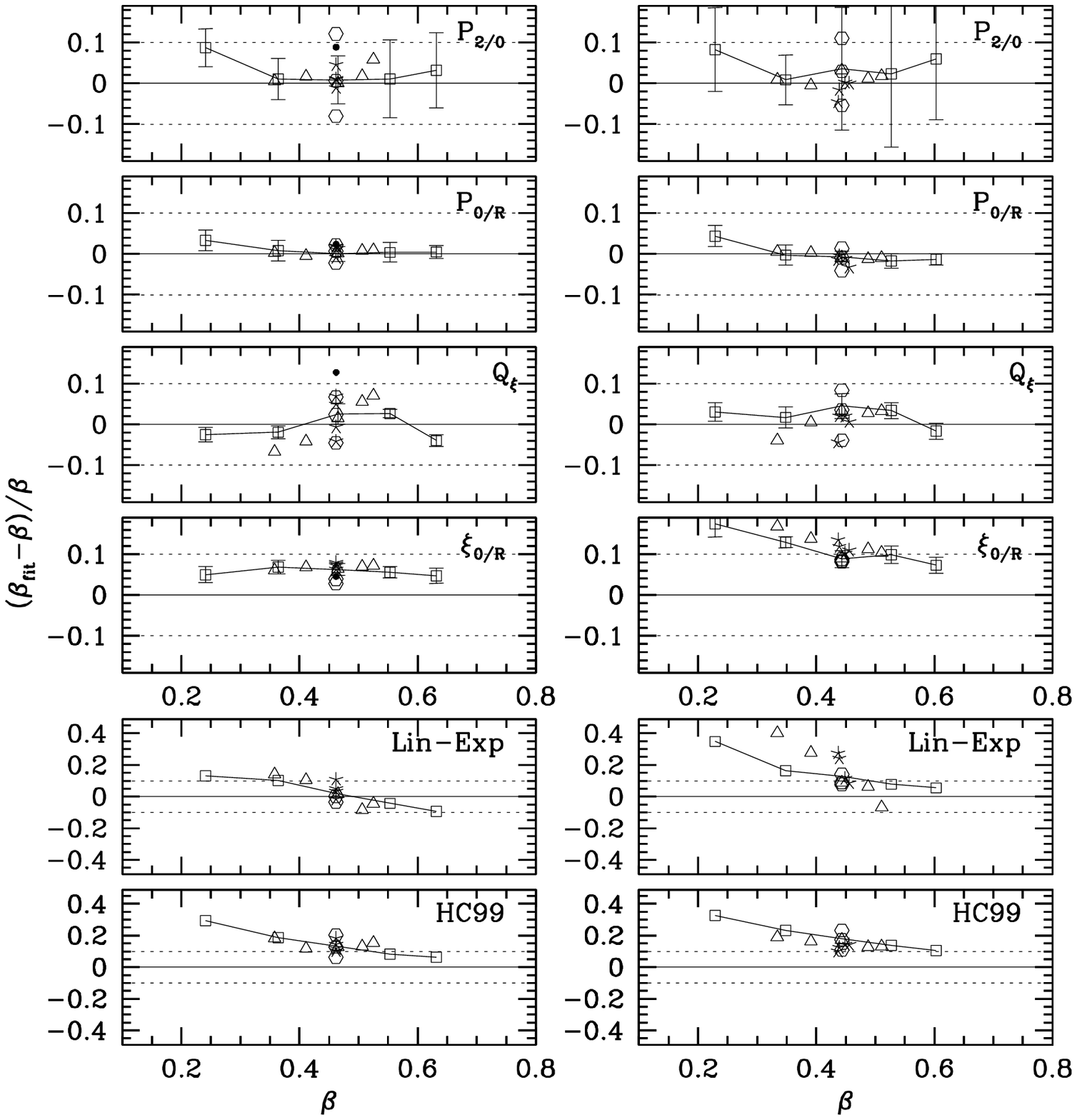,width=\two_col_fig}}
\caption{ \label{beta_errors} The relative errors in $\bfit$
  obtained from the fitting functions, for models with $\Gamma=0.2$
  (left) and $\Gamma=0.12$ (right). In each panel, the solid lines
  represent zero error, while the dotted lines are $\pm 10\%$
  error. The four different point types correspond to the four model
  sets in Table 2: constant \ma\ (squares), constant \mb\
  (triangles), constant \mc\ (stars), constant \md\ (hexagons). The
  $\av=0$ model is not shown. The $\avc=1$ model is plotted with a
  solid circle (left panels only). The bottom panels show, with an
  expanded vertical scale, the systematic errors of the
  linear-exponential model and the HC99 fitting method. }
\end{figure*}

We use a similar procedure to estimate $\beta$ from
$\pkzr(\lambda)$. Here we define the non-linear scale $\lambda_1$ as
the wavelength at which $P_Z = P_0$, and we determine it by fitting a
straight line to the six data points around $\pkzr=1$. We fit the
functional form

\begin{equation}
\label{e.fpkzr}
\pkzr (\lambda) = (\pkzrlin-1) \left[ 1 - \left(\frac{\lambda}{\lambda_1}\right) 
  ^{-1.57/(0.50+\bfit)} \right]+1,
\end{equation}

\noindent where $\bfit$ and $\pkzrlin$ are related by equation
(\ref{e.pkzr}). We estimate $\bfit$ by minimizing $\chi^2$ for all
data points with $\lambda \ge \lambda_1$. As with equation
(\ref{e.fpkqp}), the form of the exponent captures our numerical
finding that higher $\beta$ models approach asymptotic behavior more
slowly. In this case, we found that using $\bfit$ rather than
$\pkzrlin$ in the denominator of the exponent produced more accurate
results.

For $\xiQp$, we adopt the fitting function 

\begin{equation}
\label{e.fxiqp}
\xiQp(r) = \xiQp^{\rm lin} \left[ 1 - \left( \frac{1.45}{\beta_{\rm fit}}\right) 
  ^{0.75(1-r/R_0)} \right],
\end{equation}

\noindent where once again $\xiqplin$ is the free parameter and its
relation to $\bfit$ is defined in equation (\ref{e.xiqp}). The
parameter $R_0$ is the scale at which $\xi_2(r)=0$. Since the data for
$\xiQp$ are much smoother than those for the power spectrum
diagnostics, it is sufficient to fix $R_0$ by simple interpolation
between the two points surrounding $\xiQp=0$. We determine $\bfit$ by
minimizing $\chi^2$ for data points with $r \ge R_0$.

For $\xizr$, we find that the most effective method to estimate
$\beta$ is simply to fit a straight line to all data above $r=10$
\hmpc, and calculate $\bfit$ from linear theory. A minimum scale below
10 \hmpc\ allows non-linearities to affect the fit, while a larger
minimum scale reduces the precision because the error bars increase
monotonically with $r$.

Figure \ref{beta_errors} presents the main quantitative results of
this section, showing the fractional error $\epsilon \equiv
(\bfit-\beta)/\beta$ of the $\beta$ estimates from $\pkqp$, $\pkzr$,
$\xiQp$, and $\xizr$, using the fitting procedures described
above. For the left hand panels, we fit the curves shown in Figures
(\ref{pk_zr})---(\ref{xi_qp}), which are averaged over three
projections of the five $\Gamma=0.2$ simulations. Right hand panels
show results of the same procedures for the $\Gamma=0.12$ simulations.

Squares represent the fixed \mb\ model sequence, with the $\om$ range
$0.1-0.5$ producing $\beta$ values from 0.24 to 0.63 (see Table
2). The fixed \ma\ sequence, shown by the triangles, spans a narrower
range of $0.36 \le \beta \le 0.53$, since we limit $\s8$ to the range
$0.6-0.95$. Five-point stars represent \mc\ models, which all have
$\beta=0.46$ by construction. Hexagons represent the $\av=0.8$ and
$\av=1.2$ models from the fixed \md\ sequence. The $\av=1.0$ model is
the same as the central model by definition, and the model with
$\avc=0.2$ is indistinguishable from it in practice, so we omit it
from the plot. The $\avc=1$ model is shown with the small filled
circle (left panels only). We do not show results for the FOG
compression model because our fitting procedures do not apply to it.


\begin{table}
  \caption{ Errors in $\bfit$}
  \begin{tabular}{@{}ccccc@{}}
  \hline

  & \multicolumn{2}{c}{$\Gamma=0.2$} & 
  \multicolumn{2}{c}{$\Gamma=0.12$}\\  \hline

  Method & 
  $\langle \epsilon^2 \rangle^{1/2}$ [\%] &
  $\langle \epsilon \rangle$ [\%] &
  $\langle \epsilon^2 \rangle^{1/2}$ [\%] &
  $\langle \epsilon \rangle$ [\%] \\

  \hline

  $\pkqp$ & 4.4 & 1.9 & 4.2 & 1.9 \\
  $\pkzr$ & 1.3 & 0.6 & 1.8 & -0.7 \\
  $\xiQp$ & 4.3 & 0.7 & 3.5 & 1.7 \\
  $\xizr$ & 6.1 & 5.9 & 11.3 & 10.9 \\
  Lin+Exp & 9.4 & 4.2 & 18.4 & 14.6 \\
  HC99 & 14.9 & 14.0 & 17.3 & 16.5 \\

  \hline

  \end{tabular}

\end{table}

For the fixed \mb\ sequence, we calculate the statistical uncertainty
in our estimate of the fractional error $\epsilon$ by separately
fitting the five simulations in turn, then dividing the dispersion of
the $\bfit$ values by $\sqrt{5-1}=2$ to obtain the uncertainty in the
mean. These uncertainties are shown by error bars on the squares in
Figure \ref{beta_errors}. In many but not all cases, our measurement
of the bias in $\bfit$ for a given model is consistent with zero, or
only marginally inconsistent with it. However, even when the offsets
from zero are within the error bars, the trend with model parameters
along a sequence may be significant, since all of our models are based
on the same set of simulations. The total volume of our simulations is
$5\times (253$\hmpc$)^3$, equivalent to that of redshift survey
covering 8000 square degrees to a limiting depth of 460 \hmpc. Since
the three orthogonal projections sample different random orientations
of the large scale structures in each simulation, the effective volume
is somewhat larger, though the increase is not a full factor of three
because real-space structures are the same in each projection. The
error bars in Figure \ref{beta_errors} are therefore similar in
magnitude to the statistical error expected from the full SDSS
redshift survey, which will cover 8000 square degrees with a median
galaxy redshift $\sim 0.1$ (\citealt{strauss02}).

Table 3 summarizes the performance of the four $\beta$-estimators,
listing the mean and rms value of the fractional errors plotted in
Figure \ref{beta_errors}. Note, however, that the numbers depend on
the particular set of models we have chosen, so they are only a rough
indicator. For $\xizr$, our procedure of fitting a straight line to
the measurements above 10 \hmpc\ gives a precise but not accurate
value of $\bfit$, as seen earlier in Figure \ref{xi_zr}. The mean
offset is 5.9\% for $\Gamma=0.2$ and 10.9\% for $\Gamma=0.12$. The rms
values of $\epsilon$ are only slightly larger, consistent with the
small scatter around the the mean offset seen in Figure
\ref{beta_errors}, though for $\Gamma=0.12$ there is a weak but
clearly significant trend of $\epsilon$ with $\beta$. Increasing the
minimum fit radius above 10 \hmpc\ reduces the correlation but does
not eliminate the higher mean error.

The $\pkzr$ fits yield accurate $\beta$ estimates, with mean errors of
less than $1\%$ that are within the statistical uncertainty of our
calculations. The rms errors are only $1.3\%$ and $1.8\%$ for
$\Gamma=0.2$ and $\Gamma=0.12$, respectively. Velocity bias does have
a noticeable effect on the $\pkzr$ estimator, with $\pm 20\%$ changes
in $\av$ producing $\mp 2.4\%$ changes in $\bfit$. 

Errors for the quadrupole estimators $\pkqp$ and $\xiQp$ are larger,
in part because of our larger statistical uncertainties, but also
because of stronger variation with model parameters. Velocity bias has
a significant impact on $\pkqp$, with $\pm 20\%$ changes in $\av$
producing $\mp 9\%$ changes in $\bfit$ for $\Gamma=0.2$. For $\xiQp$
the effect is smaller, $\mp 5.5\%$. The slope traced by the triangular
points shows that the bias of the $\xiQp$ estimator changes steadily
with $\s8$, from $-6\%$ at $\s8=0.6$ to $+8\%$ at $\s8=0.95$ for
$\Gamma=0.2$. A similar trend with $\s8$ appears in the
constant-$\beta$ sequence.

For comparison, the lower panels of Figure \ref{beta_errors} show the
results of applying the HC99 and linear-exponential models to our
simulation results. The HC99 procedure is applied to $\pkqp$
measurements with $P_2(k)\ge 0$, and we implemented the
linear-exponential model by minimizing $\chi^2$ with respect to \xx\
for all data with $r\ge 5$ \hmpc. Note the larger vertical scale on
these panels. The HC99 simulations emphasized values of $\beta\ga
0.6$, and for $\beta\ga 0.5$ we also find it to be fairly
accurate, with a bias $\sim 10\%$. However, for lower $\beta$ values
the HC99 procedure substantially overestimates the true $\beta$, and
our modification defined by equation (\ref{e.fpkqp}) is a major
improvement.

The linear-exponential model performs reasonably well for
$\Gamma=0.2$, but there is a steady trend from positive bias at low
$\beta$ to negative bias at high $\beta$, and the rms error of $9.4\%$
is substantially larger than for any of our estimators. Increasing the
minimum fitting scale from 5 \hmpc\ to 10 \hmpc\ makes little
difference. For $\Gamma=0.12$ the linear-exponential model breaks down
more seriously, overestimating $\beta$ by up to 40\%, and showing
strong correlation of the $\bfit$ error with $\beta$ and with
$\s8$. 

By determining non-linear scales directly from the data, our
$\beta$-fitting procedures avoid any explicit dependence on $\s8$,
$\om$, or $\av$. Or course, for known values of $\s8$ or $\av$, one
could use Figure \ref{beta_errors} to remove the bias of the
estimator, further improving its accuracy. Our fitting formulas
(\ref{e.fpkqp})---(\ref{e.fxiqp}) are obtained empirically, with only
a qualitative relation to a full physical model. However, they
successfully describe models with a wide range of physical parameters,
and we will show in \S 4.5 below that the non-linear scales in these
fits depend on $\s8$, $\om$, and $\av$ in physically sensible ways.

The estimates based on redshift-space to real-space ratios, $\pkzr$
and $\xizr$, perform more robustly than those involving quadrupole
moments, once the linear theory estimate from $\xizr$ is corrected for
systematic bias. Furthermore, the monopole components $P_0(k)$ and
$\xi_0(r)$ can be measured with higher precision than the quadrupoles
$P_2(k)$ and $\xi_2(r)$, for a data set of fixed size. However, we
have not addressed the problem of determining the real-space
quantities $P_R(k)$ and $\xi_R(r)$. \cite{hamilton00} propose methods
for recovering the former, by combining the monopole, quadrupole, and
hexadecapole on large scales, and using the power in modes transverse
to the line of sight on small scales. For $\xi_R(r)$, one can invert
the projected correlation function \wp\ (see
\citealt{davis83,zehavi04b}). Alternatively, having fit \wp\ with an
HOD model, one can take the three-dimensional correlation function of
that model to represent \xg. It is possible that estimating $P_R(k)$
or \xg\ in these ways will degrade the performance of the
redshift-to-real space estimators, introducing systematic errors or
larger statistical errors. We leave that question to future work that
involves mock catalogs tailored to specific data sets.

\subsection{From $\beta$ and $\om$ to $\s8$}

With sufficiently good observational data, the procedures described in
\S 4.3 can provide estimates of $\beta\equiv \om^{0.6}/\bg$ that are
accurate to a few percent or better. For a specified value of $\om$,
this estimate in turn yields an estimate of $\bg$. However, for
cosmological purposes we are less interested in $\bg$ {\it per se}
than in the dark matter fluctuation amplitude $\s8$. In this paper we
define $\bg$ to be the mean value of $[\xi_g(r)/\xi_m(r)]^{1/2}$ over
the range 4 \hmpc\ $\le r \le$ 12 \hmpc, where the average is inverse
variance weighted and $\xi_m(r)$ is the non-linear correlation
function of the simulation dark matter particles. The value of $\bg$
is insensitive to increases in the inner or outer cutoff on the
averaging regions, though it drops if the minimum radius is pushed
much below 4 \hmpc. For example, changing the range to 10 \hmpc\ $\le
r \le$ 25 \hmpc, changes $\bg$ of the central model from 1.041 to
1.026, the largest change of the five models.

\begin{figure}
\centerline{\psfig{figure=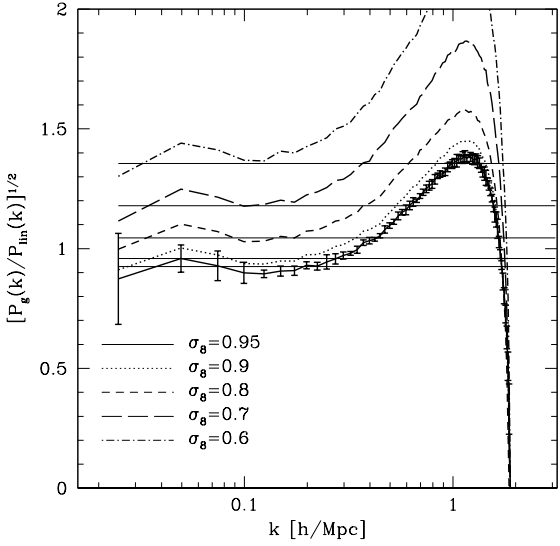,width=\one_col_fig}}
\caption{ \label{linear_bias} Comparison of bias definitions. Curves
  show the square root of the ratio of the non-linear galaxy and
  linear matter power spectra, for the five $\s8$ values as
  indicated. Error bars shown for the $\s8=0.95$ model are the error
  in the mean from five realizations. Thin lines are the value of
  $b_g$ measured from the correlation function ratio and used in our
  definition of $\beta$. }
\end{figure}

The standard analytic approximation for the large-scale bias factor,

\begin{equation}
\label{e.bg_hod}
\bg = \frac{1}{\bar{n}_g}\, \int_0^\infty\,b_h(M)\, \navg \, \frac{dn}{dM}\, dM,
\end{equation}

\noindent describes our numerical results for $\bg$ with an rms error
of 0.4\% for $\Gamma=0.2$ and 0.6\% for $\Gamma=0.12$, if we use the
halo bias formula $b_h(M)$ of \cite{tinker04} and the halo mass
function $dn/dM$ of \cite{Jenkins01}. The bias $\bg$ is a
monotonically decreasing function of $\s8$, since we match the same
galaxy correlation function by construction. The most robust way to
convert a value of $\bfit$ to a value of $\s8$ (for a specified $\om$)
is to consider a sequence of models of increasing $\s8$, carry out HOD
fits to match the observed projected correlation function \wp\ in each
case, compute $\bg$ from $\navg$ using equation (\ref{e.bg_hod}), and
pick the value of $\s8$ for which $\om^{0.6}/\bg=\bfit$.

By definition, $\s8$ is given by an integral over the linear theory
dark matter power spectrum $\plin$. In the linear approximation, where
$b_g^2 = P_g(k)/\plin$, one can use an estimated $\bg$ and the
measured galaxy power spectrum $P_g(k)$ to normalize $\plin$ and thus
compute $\s8$. Figure \ref{linear_bias} compares our definition of
$\bg$ (horizontal lines) to the power spectrum ratios
$[P_g(k)/\plin]^{1/2}$ of the $\Gamma=0.2$ simulations. For all five
values of $\s8$, the power spectrum ratios are consistent with a
constant asymptotic value at large scales, and this asymptotic value
is consistent with the value of $\bg$ defined from the correlation
function ratio. However, even with our $360^3$ simulations, we cannot
make this comparison at a precision better than a few percent because
there are relatively few Fourier modes in the asymptotic
regime. Furthermore, the power spectrum ratios lie slightly above
$\bg$ for $\s8=0.6$ and slightly below for $\s8=0.95$, with a steady
trend in between. The same trend appears in Table 1, where the product
$\s8\bg$ rises from 0.81 to 0.88 as $\s8$ grows from 0.6 to
0.95. Thus, simply normalizing $\plin$ by $P_g(k)/\bg^2$ would not
accurately describe our results at the few percent level. The trend of
$\s8\bg$ arises because we set our HOD parameters by fitting the
galaxy correlation function in the linear and non-linear regime; at
the few percent level, our large-scale galaxy correlation function is
higher for high $\s8$ (see Figure \ref{real_space_xi}). If we forced a
perfect match of the galaxy correlation function at large scales, then
$\s8\bg$ would be constant, but we could no longer match \xg\ as well at
small scales, at least with our three-parameter HOD.

The passage from $\beta$ and $\om$ to $\s8$ would be easy if we
defined the galaxy bias $\bg=b_8\equiv \sigma_{8,g}/\s8$, where
$\sigma_{8,g}$ is the (non-linear, shot noise subtracted) rms galaxy
count fluctuation in 8 \hmpc\ spheres. In this case, one could simply
divide $\bfit$ by $\om^{0.6}$ and multiply by the measured $\sigma_{8,g}$
to obtain $\s8$. We have tried to develop procedures like those in \S
4.3 to estimate $\beta_8\equiv \om^{0.6}/b_8$. However, once we tune
the estimation formulas to the $\Gamma=0.2$ simulations, they do not
provide accurate results for $\Gamma=0.12$, in contrast to our
procedures for $\beta$, which give accurate results for both power
spectrum shapes. An 8 \hmpc\ top-hat does not suppress non-linear
clustering enough for the bias factor $b_8$ to approximate bias in the
linear regime (as also noted by HC99).

In Paper II, we develop an analytic approach that circumvents the
complication of mapping $\beta$ into the $\s8 - \om^{0.6}$ parameter
space, as the fitting parameters are $(\om,\s8,\av)$, without
reference to $\beta$.

\subsection{Length Scales in Large-Scale Distortions}

\begin{figure}
\centerline{\psfig{figure=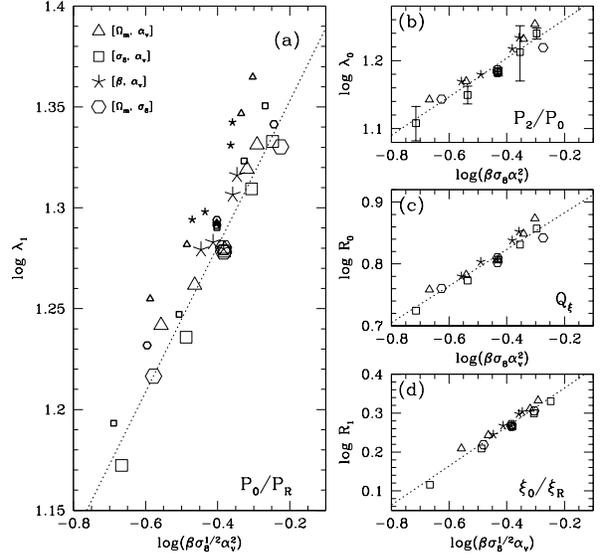,width=\one_col_fig}}
\caption{ \label{r0_4win} Influence of cosmological parameters on the
  non-linear length scales of the large-scale distortion measures. (a)
  The wavelength $\lambda_1$ at which $P_0/P_R=1$ is plotted against
  $\beta \sigma_8^{1/2} \alpha_v^2$. Large and small points show the
  $\Gamma=0.2$ and $\Gamma=0.12$ results, respectively, with point
  types corresponding to model sequences as indicated in the
  legend. (b) --- (d) Non-linear length scales for the other three
  large-scale measures, plotted against the combinations of $\beta$,
  $\sigma_8$, and $\alpha_v$ that produce the least scatter, for
  $\Gamma=0.2$. Dotted lines show least squares fits to the data
  points. }
\end{figure}

The distortions in redshift-to-real space and quadrupole-to-monopole
ratios in Figures \ref{pk_zr} and \ref{xi_qp} are driven mainly by
galaxy velocity dispersions on small and intermediate scales, which
drive down the redshift-space correlation amplitude and reverse the
sign of quadrupole distortions. The non-linear length scales in
equations (\ref{e.pkzr}), (\ref{e.pkqp}), and (\ref{e.xiqp}), and the
radius $R_1$ at which $\xi_0 = \xi_R$, therefore encode information
about the parameters $\om$, $\s8$, and $\av$, as an increase in any of
these variables increases the galaxy velocity dispersion.  The
dependence of the galaxy velocity dispersion on $\om$ is
straightforward: at fixed $\s8$, the large-scale velocity field
follows the linear theory scaling $\om^{0.6}$, and the virial
velocities of halos of fixed abundance scale as $\om^{1/2}$ (ignoring
the small dependence of halo concentration on $\om$).  These two
effects appear at different scales, but we find that the pairwise
galaxy velocity dispersion scales roughly as $\om^{0.55}$ in our
simulations at all separations. For $\s8$ and $\av$, the situation is
more complicated. Velocity bias is most influential at small scales,
where the galaxy pairs come from within one halo. At larger scales, a
significant fraction of pairs involve the central galaxies of low-mass
halos, and are thus not affected by satellite velocity
bias. Inspection of our numerical results suggests that at large
separations the pairwise dispersion scales as $\av^{1/2}$. The power
spectrum normalization affects the galaxy velocity dispersion in two
ways: at linear scales the halo velocity dispersion increases linearly
with $\s8$, while the internal velocity dispersions of halos hosting
multiple galaxies increase with $\s8$ because of the higher halo
masses.

Inspection of the analytic solution for $\pkzr$ in the
linear-exponential model (see \citealt{cfw95}, \S 2.1) implies that
the non-linear scale $\lambda_1$ where $\pkzr=1$ should scale linearly
with the velocity dispersion $\sigma_v$ at fixed $\beta$ and
approximately as $\beta^{-1/2}$ at fixed $\sigma_v$. With the scalings
$\sigma_v\propto \om^{0.55}\av\s8$ discussed above, we obtain

\begin{equation}
\label{e.lam1}
\lambda_1 \propto \beta^{-1/2} \sigma_v \propto \om^{0.25} \s8^{0.5} \av,
\end{equation}

\noindent where the last relation uses $\beta \propto
\s8\om^{0.6}$.

The left-hand panel of Figure \ref{r0_4win} plots $\lambda_1$ against
$\beta \sigma_8^{1/2} \av^2$, a combination of parameters chosen by
trial and error to yield minimal scatter. The numerical data form a
tight power-law for the $\Gamma=0.2$ models. The statistical errors
derived from the run-to-run dispersion are of order the point size,
and the fit has a $\chi^2$ per degree of freedom of 8.9, indicating
that most of the model-to-model scatter is physical in origin. The
data for the $\Gamma=0.12$ models follow the same slope, but the
amplitude of the relation is 5\% higher, and there is more
scatter. The dotted line plotted in the panel is a least squares fit
to the $\Gamma=0.2$ data. The slope is $\sim 0.36$, making a scaling
of $\lambda_1 = 26.3\, \om^{0.22} \s8^{0.54} \av^{0.72}$ \hmpc. Given
the approximate nature of the arguments behind equation
(\ref{e.lam1}), the agreement with the numerically derived scaling is
quite good. The lower index on $\av$ in the numerical results arises
because the scale $\lambda_1 \sim 10$ \hmpc\ is outside the
one-halo regime where $\sigma_v \propto \av$ but not fully in the
large scale regime where $\sigma_v \propto \av^{1/2}$.

The remaining panels of Figure \ref{r0_4win} plot the other non-linear
length scales against a combination of parameters chosen by trial and
error to produce minimum scatter. For the central model, the $\pkqp$
zero-crossing $\lambda_0$ is slightly smaller than $\lambda_1$,
$\lambda_0$ is $\sim 2$ times the $\xiQp$ zero-crossing $R_0$, and
$R_0$ is $\sim 3$ times the scale $R_1$ at which $\xizr=1$. Dotted
lines show best-fit power-law relations, $\lambda_0 =
20.7\,(\beta\s8\av^2)^{0.28}$, $R_0 = 8.7(\beta\s8\av^2)^{0.30}$, $R_1
= 2.9\,(\beta\s8^{1/2}\av)^{0.50}$. Scatter for the quadrupole length
scales is consistent with the statistical errors (see panel b), which
are larger for these measurements.

In principle, these non-linear length scales can help determine
cosmological parameters by adding another observable quantity to break
degeneracies in our three-dimensional parameter space. For example,
once $\beta$ is fixed by the large-scale distortions, the measurement
of $\lambda_1$ constrains the parameter combination
$\s8^{0.18}\av^{0.72}$. Since the different length scales have
different parameter dependencies, once can use combinations to isolate
$\av$ and $\s8$. For example, the best-fit power laws imply

\begin{equation}
\label{e.av_scalelength}
\av = \left( \frac{\lambda_1}{26.9 \,h^{-1}\mbox{Mpc}}\right)^{2.78}
\left( \frac{R_1}{2.92\,h^{-1}\mbox{Mpc}}\right)^{-2.00}.
\end{equation}

\noindent The models with no velocity bias ($\av=1$) follow this
relation with an rms error of 3.8\% and a mean error of $-2.4\%$. For
the models with $\av=0.8,1.2$, equation (\ref{e.av_scalelength})
predicts 0.80 and 1.10 respectively. The power law fits for $R_0$ and
$\lambda_1$ yield

\begin{equation}
\label{e.fit_s8}
\sigma_8 = \left( \frac{R_0}{8.71\,h^{-1}\mbox{Mpc}}\right)^{6.78}  
\left( \frac{\lambda_1}{26.9\,h^{-1}\mbox{Mpc}}\right)^{-5.56}.
\end{equation}

\noindent The values of $\s8$ predicted with equation
(\ref{e.fit_s8}) are accurate to within an rms error of 12.6\%.


\begin{figure*}
\centerline{\psfig{figure=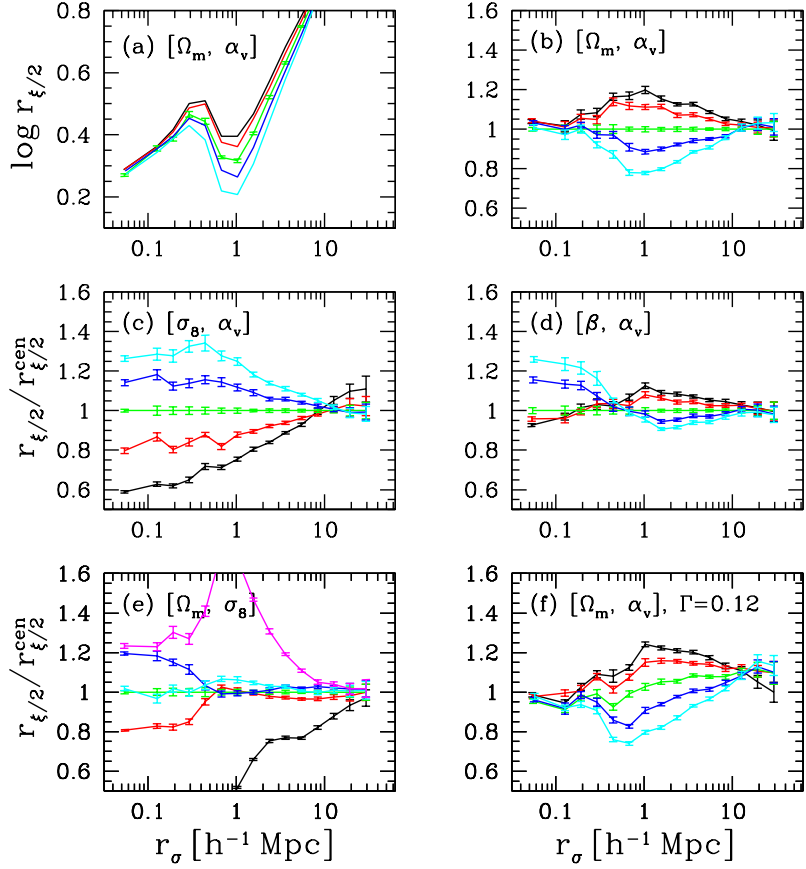,width=\two_col_fig}}
\caption{ \label{rhalf_sigma} The smalle-scale distortion parameter,
  $\rhalf$, as a function of transverse separation. For panels (b) ---
  (f), each curve has been normalized by the values for the central
  model, while panel (a) plots the curves as $\log\,\rhalf$ for the
  constant \ma\ sequence, without normalization. Models run from
  $\s8=0.95$ ({\it black}) to $\s8=0.6$ ({\it cyan}) in panels (a) and
  (b), from $\om=0.1$ ({\it black}) to $\om=0.5$ ({\it cyan}) in panel
  (c), from $\s8=0.95$ ({\it black}) to $\s8=0.6$ ({\it cyan}) in
  panel (d), with the order of the colors being black, red, green,
  blue, cyan. In panel (e), models are $\av=0$ ({\it black}),
  $\av=0.8$ ({\it red}), $\av=1.0$ ({\it green}), $\av=1.2$ ({\it
  blue}), $\avc=0.2$ ({\it cyan}), and $\avc=1$ ({\it magenta}). (f)
  Models are the same as panel (b), but with $\Gamma=0.12$. }
\end{figure*}

\section{Small-Scale Distortion}

While the non-linear length scales give some measure of small-scale
velocities, we can characterize these velocities more physically and
more accurately by focusing on distortions at small \rsig, where they
dominate. The traditional measure of small-scale distortions is the
pairwise velocity dispersion, but this is not a direct observable; it
is extracted from the data by fitting a model that specifies the scale
dependence of the mean pairwise velocity of galaxies and the form of
the velocity distribution (e.g., Davis \& Peebles 1983). We would
prefer a quantity that is measured directly from the data, and here
we follow the lead of Fisher \etal\ (1994), who use $\xi(r_\pi)$ at
fixed, small \rsig. Referring back to Figure \ref{sigma_pi}, we see
that $\xi(r_\pi)$ at small \rsig\ is constant for a range of \rpi,
before turning over at a scale determined by the galaxy velocity
dispersion. We can quantify this turnover by the measure $\rhalf$, the
value of \rpi\ at which the correlation function decreases by a factor
of two relative to its value at $r_\pi=0$. More generally, one could
use the shape of $\xi(r_\pi)/\xi(r_\pi=0)$ over some range of the
line-of-sight separation, scaling by $\xi(r_\pi=0)$ to remove the
sensitivity of the distortion measure to the exact value of the
real-space correlation function.

Figure \ref{rhalf_sigma}a plots $\rhalf$ against \rsig\ for the \ma\
sequence. All the curves have a characteristic wave pattern, which
rises to a maximum at $r_\sigma \sim 0.6$ \hmpc\ and reaches a minimum
at $r_\sigma\sim 1$ \hmpc. The rise at small separation is the result
of including one-halo galaxy pairs from increasingly more massive
halos with higher velocity dispersions. The minimum at 1 \hmpc\ occurs
near the one-halo to two-halo transition in the real-space \xg. At
this separation, two-halo pairs come largely from the central galaxies
of lower mass halos, so they do not have an internal dispersion
contribution, and the halo pairwise velocities themselves are
relatively low. At $r_\sigma > 1$ \hmpc, all curves monotonically
increase, as the internal dispersions of large halos again start to
contribute and the pairwise dispersion of halos themselves
increases. To highlight the differences between the models, panels (b)
--- (f) plot five model sequences where all the curves have been
normalized by the values for the central model
$(\om=0.3,\s8=0.8,\av=1,\avc=0)$. Panels (b) --- (e) show the standard
suite from Table 2 and earlier figures. In panel (b), with fixed $\om$
and $\av$, changing $\s8$ has little effect on $\rhalf$ at
$r_\sigma\le 0.2$ \hmpc. This separation is small enough that rare,
high-mass halos do not contribute a large fraction of the one-halo
galaxy pairs relative to the pairs contributed by halos with mass $M_h
\approx M_1$, where $\s8$ has little effect on the halo mass function.
The value of $\s8$ has a large impact on $\rhalf$ at $r_\sigma \sim 1$
\hmpc, the location of the one-halo to two-halo transition. More high
mass halos create more large separation one-halo pairs, extending the
one-halo \xg\ to larger $r$. These pairs have large velocity
dispersion and are therefore spread out along the line of sight,
increasing $\rhalf$.

In panel (c), with fixed $\s8$ and $\av$, changing $\om$ affects
$\rhalf$ at all $r_\sigma\la 10$ \hmpc. Higher $\om$ increases
both halo pairwise velocities and internal velocity dispersions, thus
increasing $\rhalf$ on all scales where dispersion dominates over
coherent flows.  Panel (d) shows models with constant $\beta$ and
$\av$, and thus constant large-scale anisotropy. As expected from the
previous results, higher $\om$ models have larger $\rhalf$ at
$r_\sigma\sim 0.1$ \hmpc, where $\s8$ has little impact. At
$r_\sigma\sim 1$ \hmpc, the higher $\om$ models (with lower
$\s8$) have smaller $\rhalf$; the depression seen in panel (b) wins
out over the enhancement in panel (c).  Thus, at fixed $\beta$ and
$\av$, the small scale distortions can break the degeneracy between
$\om$ and $\s8$.

Panel (e) shows models with varying $\av$ but constant $\om$ and $\s8$
(and thus constant $\beta$). Not surprisingly, the $\av=0$ model has
very small values of $\rhalf$ relative to the central model at scales
less than 10 \hmpc. The effect of moderate velocity bias is most
significant at the smallest \rsig, with 20\% changes in $\rhalf$ at
$r_\sigma=0.1$ \hmpc\ for $\av=1.2$ or 0.8. However, these $\av$
variations have little impact at large \rsig, where 20\% changes of
internal velocity dispersions are small compared to halo velocities
themselves, and the effect is essentially zero at $r_\sigma\sim 1$
\hmpc. At this separation, two-halo pairs begin to dominate \xx, but
\rsig\ is still smaller than the virial radii of large halos. Most
pairs therefore come from halos that contain a central galaxy and no
satellites, and the value of $\av$ has no effect. Central galaxy
velocities have maximum effect at the $\sim 1$ \hmpc\ scale, for the
same reason. Setting $\avc=0.2$ boosts $\rhalf$ by 5-10\% at this
\rsig, while treating central galaxies like satellites ($\avc=1$)
boosts it by a factor of two.

Panel (f) plots the results for the constant \ma\ sequence with
$\Gamma=0.12$, once again normalized by the $\Gamma=0.2$ central
model. As in panel (b), $\s8$ has minimal effect at small scales and
makes the most difference at $r_\sigma\sim 1-2$ \hmpc. The higher
$\rhalf$ at large \rsig\ in the $\Gamma=0.12$ models probably reflects
the shallower real-space correlation function at these scales.

\begin{figure*}
\centerline{\psfig{figure=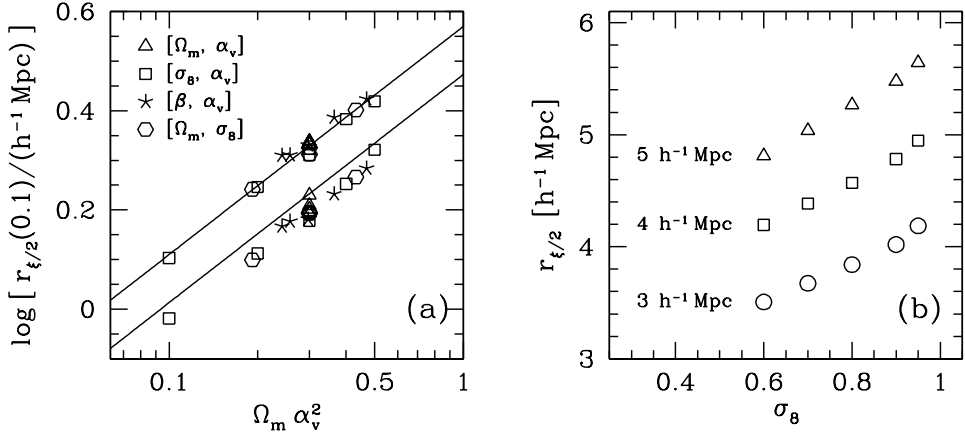,width=\two_col_fig}}
\caption{ \label{rhalf_0.1and3} Diagnostic power of the small scale
  distortion measure. (a) Points show $\rhalfa$ against $\om\av^2$ for
  all of the $\Gamma=0.2$ models, except $\av=0$ and $\avc=1$. The
  upper line shows a least-squares fit, $\rhalfa = 3.7(\Omega_m
  \alpha_v^2)^{0.46}$, with $\chi^2_{d.o.f.}=1.9$. The lower
  points, shifted down by 0.2 dex for visual clarity, show the
  $\Gamma=0.12$ results. These data lie $\sim 7\%$ below the (shifted)
  line. (b) Dependence of $\rhalf$ on $\s8$ for models in which both
  the large scale distortions ($\beta=0.46$) and the small-scale
  distortions [$\rhalfa=2.0$ \hmpc] are fixed to the same
  values. Circles, squares, and triangles represent $r_\sigma = 3, 4$
  and 5 \hmpc, respectively. }
\end{figure*}

Figure \ref{rhalf_sigma} demonstrates that $\rhalf$ is a robust
diagnostic for $\om$ and $\av$ when \rsig\ is small, independent of
$\s8$ or $\Gamma$. In figure \ref{rhalf_0.1and3}a, the upper points
plot $\rhalfa$ against $\om\av^2$ for all of the $\Gamma=0.2$ models
(except those with $\av=0$ and $\avc=1$). The data follow a power law
with a slope of 0.46 and minimal scatter. For one-halo pairs, the
redshift-space separation depends on relative velocities, which are
proportional to $\om^{1/2} \av$, and one might therefore expect a
slope of 0.5. Because there is a small two-halo contribution to \xx\
at these separations, the slope deviates slightly from this
expectation. The data for $\Gamma=0.12$ follow a similar power law,
but with a normalization $\sim 7\%$ lower, as expected from the results
in Figure \ref{rhalf_sigma}f. This offset may arise partly from the
difference in the real-space correlation function, which is shallower
for $\Gamma=0.12$, and partly from the difference in the halo mass
function, which changes the relative importance of pairs from
different halos.

The values of $\beta$ and $\rhalfa$ provide two observable constraints
in our three-dimensional $(\om,\s8,\av)$ parameter space, measuring
the combinations $\s8 \om^{0.6}$ and $\om \av^2$. A measurement of
$\rhalf$ at somewhat larger \rsig\ has the possibility of providing a
third constraint on a different combination of these parameters. Based
on the power-law fit in Figure \ref{rhalf_0.1and3}a, each
constant-$\beta$ model was given the value of $\av$ required to match
$\rhalfa$ of the central model. Relative to Figure \ref{rhalf_sigma}d
at fixed \mc, this scaling brings curves together at $r_\sigma< 1$
\hmpc, but it makes little difference at larger separations where
$\av$ has little effect.  Differentiating between adjacent models
requires high precision in the measurements, but there is a clear,
20\% separation between the low and high values of $\s8$ with this
diagnostic. In Figure \ref{rhalf_0.1and3}b, we plot $\rhalf$ against
$\s8$ for $r_\sigma=3, 4$, and 5 \hmpc. At each transverse separation,
there is a monotonic, nearly linear trend with $\s8$ once $\beta$ and
$\rhalfa$ have been fixed. These results allow for unambiguous
determination of $\s8$, breaking the third and last degeneracy in the
parameter space.

Figure \ref{rhalf_0.1and3}b assumes $\avc=0$, and central galaxy
velocities could interfere with this approach to breaking
degeneracies. For example, adopting $\avc=0.2$ increases $\rhalfa$ by
$\sim 5\%$, which is of order the effect of changing $\s8$ by
0.1. However, the effects of moderate $\avc$ on this measure go away
at scales larger than 3 \hmpc, where there is still clear model
differentiation in Figures \ref{rhalf_sigma}d and
\ref{rhalf_0.1and3}b. As we have already noted, physical arguments and
hydrodynamic simulations support the assumption of low $\avc$, but
further theoretical and observational investigation of this point is
warranted.

\section{Discussion}

Our results provide a blueprint for obtaining constraints in the
$(\om,\s8,\av)$ parameter space from measurements of clustering
anisotropy in redshift space. For each model in the parameter space,
one first chooses HOD parameters to reproduce measurements of the
projected galaxy correlation function \wp, which depends only on the
real-space correlation function \xg. If the assumed power spectrum
shape is correct, it will generally be possible to match \wp\ well for
a wide range of $\s8$ and $\om$. At large scales, the anisotropy
ratios $\pkzr$, $\pkqp$, $\xizr$ or $\xiQp$ then depend on
$\beta\equiv \om^{0.6}/b_g$, where $b_g^2=\xi_g/\xi_m$ is a
monotonically decreasing function of $\s8$ for fixed galaxy clustering
(see \S 4.4). These measures scale with cosmological parameters as
predicted by linear theory and the linear bias model
(\citealt{kaiser87}), even though these approximations do not provide
an accurate description of anisotropy on most scales accessible to
observations or to our simulations. One can estimate $\beta$ by fitting
$\pkzr$, $\pkqp$, or $\xiQp$ as a function of scale using our
equations (\ref{e.fpkqp}), (\ref{e.fpkzr}), and (\ref{e.fxiqp}), or by
measuring $\xizr$ at $r \ge 10$ \hmpc\ and correcting for the $\sim
6\%$ bias of linear theory (see Figure \ref{beta_errors}). The
turnover scales in the fitting functions depend on the velocity bias
$\av$, but they can be measured directly from the anisotropy ratios,
so the $\beta$ estimates themselves are largely independent of $\av$.

The turnover scales can be used to break degeneracies in the parameter
space, but the line-of-sight correlation function $\xi(r_\pi)$ at
fixed, small \rsig\ provides a more direct measure of velocity
distortions in the highly non-linear regime. In particular, for small
\rsig\ the scale $\rhalf$ defined by $\xi(r_\sigma,r_{\xi/2}) =
0.5\times \xi(r_\sigma,0)$, quantifies the typical length of
``fingers-of-god,'' and hence the characteristic amplitude of pairwise
velocity dispersions. At $r_\sigma \sim 0.1$ \hmpc, where most pairs
come from intermediate mass halos, we find that $\rhalf$ depends on
$\om \av^2$ with essentially no dependence on $\s8$. At $r_\sigma \sim
1-5$ \hmpc, $\rhalf$ has a significant dependence on $\s8$ even at
fixed $\beta$ and $\av$, with $\Delta \s8 \sim 0.1$ corresponding to
$\Delta \rhalf \sim 5\%$. Therefore, one can in principle use
measurements of large-scale anisotropy and $\xi(r_\pi)$ at $r_\sigma
\sim 0.1-5$ \hmpc\ to separately determine the values of $\om$, $\s8$,
and $\av$. Alternatively, one can measure $\beta$ and $\om\av^2$ as
described above and adopt theoretical priors on $\av$ from
hydrodynamic simulations of galaxy formation (e.g., Berlind \etal\
2003), or combine redshift-space distortions with other observables
that constrain different combinations of $\s8$ and $\om$. For example,
galaxy-galaxy lensing measurements constrain $\s8 \om$ (instead of
$\s8 \om^{0.6}$) from the ratio of the galaxy-mass correlation
function to the galaxy autocorrelation function (Sheldon \etal\
2004). The galaxy bispectrum can yield a direct estimate of $\s8$ by
determining the large-scale galaxy bias factor (Fry 1994; Verde \etal\
2002).

Our blueprint has significant advantages relative to the
linear-exponential model or the alternative fitting procedure of
HC99. First, our approach is more accurate for a wide range of
cosmological models (Figs. \ref{kaiser_compare},
\ref{beta_errors}). Averaging over both values of $\Gamma$ used, our
fitting function for $\pkzr$ yields $\beta$ with an rms error of 1.6\%
for the range of models presented. For the $\pkqp$ and $\xiQp$
diagnostics, the fitting functions yield rms errors of 4.1\% and 3.9\%
respectively. Second, our approach makes use of the small-scale
anisotropy as a tool for breaking parameter degeneracies, instead of
treating the galaxy dispersion as a nuisance parameter. Constraints on
$\s8$ and $\av$ from these small scale measures can be used to further
improve the $\beta$ estimate.

The fitting formulas presented here are designed to allow
straightforward parameter estimation given measurements of \xx\ and
$P_Z(k,\mu)$. Alternatively, one can use simulations to calibrate a
fully analytic description of redshift-space anisotropy, in which case
one can fit data directly using $\om$, $\s8$, and $\av$ as the fitting
parameters. We will develop such a model in Paper II; achieving the
accuracy demanded by data sets like the SDSS and the 2dFGRS is not
easy, but it is possible. The analytic method is more flexible than
the fitting formula approach, allowing one to take more complete
advantage of information in \xx\ or $P_Z(k,\mu)$. At the opposite
extreme, one can circumvent analytic formulations entirely and fit
data by directly populating halos of N-body simulations and measuring
anisotropy, using the $\om$-scaling technique of this paper to improve
efficiency. With large volume simulations that resolve the necessary
halo masses, this method should achieve the highest accuracy because
it fully describes non-linear halo clustering, and it can address
corrections to the distant-observer approximation and other technical
issues that are difficult to model analytically. In practice, it will
probably be best to use the fitting formulas or an analytic model to
locate the most interesting regions of parameter space, then use
focused numerical simulations to check and refine estimates.

For the $\av=0$ model, large-scale anisotropy measures agree
reasonably with linear theory over a substantial range in scale. This
result suggests that FOG compression plus linear theory is a viable
alternative approach to estimating $\beta$. Assessing the systematic
uncertainties of this method requires tests with realistic mock
catalogs that quantify the ability of the FOG compression algorithm to
correctly identify and compress true FOGs in galaxy survey data.

There are several limitations to our blueprint. With two exceptions,
we have assumed that central galaxies move with the center of mass of
the halo, i.e.\ $\avc=0$. Changing $\avc$ to 0.2 makes minimal
difference in both the large scale measures and $\rhalf$. However,
setting $\avc \approx 1$ makes a considerable difference. Current
hydrodynamical simulations (Berlind \etal\ 2003) suggest $\avc
\la 0.2$ is a reasonable assumption, but the issue merits further
investigation because of its significant impact on redshift-space
anisotropy modeling. Analysis of SDSS galaxies shows that
central-satellite galaxy pairs indeed have a narrower velocity spread
than satellite-satellite pairs (T. McKay \etal , in preparation). We have
also assumed that $\av$ is independent of halo mass. This assumption
should be adequate because most one-halo pairs come from a limited
range of halo masses; low-mass halos have no satellites, and high-mass
halos are rare. To significantly alter our results, $\av$ would need
to depend strongly on mass in the relatively narrow range $M_1/2 - 5
M_1$, and even then its effect might be well represented by an average
value. The weak mass dependence seen in the simulations analyzed by
Berlind \etal\ (2003) does not affect the results here, but the
question again merits investigation in future hydrodynamic studies of
galaxy formation. One can also test for mass dependence of $\av$ by
comparing the predicted and observed scalings of group velocity
dispersions with group richness.

The experiments illustrated in Figure \ref{hod_test} show that
changing the details of the HOD, or the assumption about spatial bias
within halos, has negligible impact on redshift-space distortions
provided one matches the same real-space clustering. However, our
investigation of these points is not exhaustive. Effects of changing
$\avc$, making $\av$ mass-dependent, or changing HOD prescriptions
while maintaining \xg\ can all be examined in more detail using the
analytic model of Paper II.

The simulations presented in this work have less dynamic range than is
ideal. At the largest scales, our numerical predictions are less
precise than the measurement precision achievable with 2dFGRS or SDSS
data, though not by a large factor. We have focused on predictions for
luminosity-threshold galaxy samples with space density $5.6\times
10^{-3}\, ($\hmpc $)^{-3}$, corresponding roughly to $M_r < -20+5\log
h$. To make predictions or test fitting formulas for fainter galaxies,
which occupy less massive halos, one would need higher resolution
simulations but similar simulation volumes. To get precise results for
more luminous galaxies that reside in rare, massive halos, one would
need larger simulation volumes, though the mass resolution required is
lower. The analytic model described in Paper II can easily be applied
to samples with different luminosity or color selection and
correspondingly different HODs, and it automatically extends to large
scales. However, additional simulations will be needed to test the
accuracy of the analytic model in these regimes.

The monopole-to-real space ratios, $\pkzr$ and $\xizr$, have smaller
systematic errors as estimators of $\beta$ than the
quadrupole-to-monopole ratios $\pkqp$ and $\xiQp$. However, we have
not addressed the problem of estimating \xg\ or $P_R(k)$ from
data. Techniques for estimating these quantities exist (e.g., Davis \&
Peebles 1983; Hamilton \etal\ 2000; Zehavi \etal\ 2004b; Tegmark
\etal\ 2004a), but we do not yet know whether they are accurate at the
$\sim 1-2\%$ level required if they are not to contribute
significantly to uncertainties in the estimates of $\beta$.

Despite these limitations, our results demonstrate that HOD modeling
can substantially improve the accuracy and precision of redshift-space
distortion analysis by replacing {\it ad hoc} extensions of linear
perturbation theory with a complete, fully non-linear description of
dark matter dynamics and galaxy bias. This level of sophistication is
required to take full advantage of data provided by the 2dFGRS and
SDSS. Precise cosmological constraints from galaxy clustering
complement those from other cosmological observables like CMB
anisotropy, gravitational lensing, the Lyman-$\alpha$ forest, or Type
Ia supernovae. They thus enhance the opportunity to detect departures
from the simplest \lcdm\ model, which could provide insight into the
physics of dark energy or the origin of primordial fluctuations.

\section*{Acknowledgments}

We thank Andreas Berlind and Roman Scoccimarro for helpful
discussions. We thank Vijay Narayanan for use of his PM code, Andreas
Berlind for code to compute the redshift-space power spectrum, and
Volker Springel for providing the public \gad\ code.  The simulations
were performed on the Beowulf and Itanium clusters at the Ohio
Supercomputing Center under grants PAS0825 and PAS0023.  ZZ
acknowledges the support of NASA through Hubble Fellowship grant
HF-01181.01-A awarded by the Space Telescope Science Institute, which
is operated by the Association of Universities for Research in
Astronomy, Inc., for NASA, under contract NAS 5-26555. JLT acknowledges
the support of a Distinguished University Fellowship at the Ohio State
University. This work was also supported by NSF grants AST-0098584 and
AST-0407125.



\begin{thebibliography}{}

\bibitem[Abazajian et al.(2004)]{kev04}
Abazajian, A.,  \etal\ ApJ (submitted), astro-ph/0408003

\bibitem[Bean et al.(1983)]{bean83}
Bean, A.\ J., Ellis, R.\ S., Shanks, T., Efstathiou, G., \& Peterson, B.\ A. 1983, MNRAS, 205, 605

\bibitem[Benson(2001)]{benson01}
Benson, A.\ J. 2001, MNRAS, 325, 1039

\bibitem[Benson et al.(2000)]{benson00}
Benson, A.\ J., Cole, S., Frenk, C.\ S., Baugh, C.\ M., \& Lacey, C.\ G. 2000, MRNAS, 311, 793

\bibitem[Berlind \& Weinberg(2002)]{bw02}
Berlind, A.\ A., \& Weinberg, D.\ H.\ 2002, ApJ, 575, 587

\bibitem[Berlind et al.(2003)]{berlind03} 
Berlind, A.\ A., Weinberg, D.\ H., Benson, A.\ J., Baugh, C.\ M.,
Cole, S., Dav\'e, R., Frenk, C.\ S., Katz, A., \& Lacey, C.\ G. 2003,
ApJ, 593, 1

\bibitem[Berlind et al.(2001)]{berlind01}
Berlind, A.\ A., Narayanan, V.\ K., \& Weinberg, D.\ H. 2001, ApJ, 549, 688

\bibitem[Blanton et al.(2003)]{blanton03}
Blanton, M.\ R., et al. 2003, ApJ, 592, 819

\bibitem[Bullock et al.(2001)]{bullock01}
Bullock, J.\ S., Kolatt, T.\ S., Sigad, Y., Somerville, R.\ S., Klypin, A.\ A.,
Primack, J.\ R., Dekel, A.\ 2001, MNRAS, 321, 559

\bibitem[Colless et al.(2001)]{colless01}
Colless, M. et. al. 2001, MNRAS, 328, 1039

\bibitem[Cole et al.(1994)]{cfw94}
Cole, S., Fisher, K.\ B., \& Weinberg, D.\ H. 1994, MNRAS, 267, 785

\bibitem[Cole et al.(1995)]{cfw95}
Cole, S., Fisher, K.\ B., \& Weinberg, D.\ H. 1995, MNRAS, 275, 575

\bibitem[Cooray (2004)]{cooray04}
Cooray, A., 2004, MNRAS, 348, 250

\bibitem[Cooray \& Sheth(2002)]{cooraysheth02}
Cooray, A. \& Sheth, R.\ K. 2002, Phys. Rep., 372, 1

\bibitem[Davis et al.(1978)]{davis78}
Davis, M., Geller, M.\ J., \& Huchra, J. 1978, ApJ, 221, 1

\bibitem[Davis \& Peebles(1983)]{davis83}
Davis, M. \& Peebles, P.\ J.\ E. 1983, ApJ, 267, 465

\bibitem[Davis et al.(1985)]{davis85}
Davis, M., Efstathiou, G., Frenk, C.\ S., \& White, S.\ D.\ M. 1985, ApJ, 292, 371

\bibitem[Efstathiou et al.(1992)]{ebw92}
Efstathiou, G., Bond, J.\ R., \& White, S.\ D.\ M. 1992, MNRAS, 258, 1

\bibitem[Faltenbacher et al.(2004)]{faltenbacher04}
Faltenbacher, A., Kravtsov, A.\ V., Nagai, D., \& Gottl\"ober,
S. 2004, MNRAS (submitted), astro-ph/0408488

\bibitem[Fisher et al.(1994)]{fisher94}
Fisher, K.\ B., Davis, M., Strauss, M.\ A., Yahil, A., \& Huchra, J.\ P. 1994, MNRAS, 267, 927

\bibitem[Fisher \& Nusser(1996)]{fisher96}
Fisher, K.\ B. \& Nusser, A 1996, MNRAS, 279, L1

\bibitem[Fry(1994)]{fry94}
Fry, J.\ N. 1994, Phys. Rev. Lett., 73, 215

\bibitem[Hamilton(1992)]{hamilton92}
Hamilton, A.\ J.\ S. 1992, ApJ, 385, L5

\bibitem[Hamilton(1998)]{hamilton98}
Hamilton, A.\ J.\ S. 1998, in The Evolving Universe, ed. D. Hamilton (Dordrecht: Kluwer), 185

\bibitem[Hamilton et al.(2000)]{hamilton00}
Hamilton, A.\ J.\ S., Tegmark, M., \& Padmanabhan, N. 2000, MNRAS, L317, 23

\bibitem[Hatton \& Cole(1998)]{hc98}
Hatton, S. \& Cole, S. 1998, MNRAS, 296, 10

\bibitem[Hatton \& Cole(1999)]{hc99}
Hatton, S. \& Cole, S. 1999, MNRAS, 310, 1137

\bibitem[Hawkins et al.(2003)]{hawkins03}
Hawkins, E., \etal\ 2003, MNRAS, 346, 78

\bibitem[Hu \& Kravtsov(2003)]{hukravtsov03}
Hu, W. \& Kravtsov, A.\ V., 2003, ApJ, 584, 702

\bibitem[Jing et al.(1998)]{jingmoborner98}
Jing, Y.\ P., Mo, H.\ J., \& B\"orner, G. 1998, ApJ, 494, 1

\bibitem[Jenkins et al.(2001)]{Jenkins01} 
Jenkins, A., Frenk, C.\ S., White, S.\ D.\ M., Colberg, J.\ M., Cole, S.,
Evrard, A.\ E., Couchman, H.\ M.\ P., \& Yoshida, N.\ 2001, MNRAS, 321, 372

\bibitem[Kaiser(1987)]{kaiser87}
Kaiser, N. 1987, MNRAS, 227, 1

\bibitem[Kang et al.(2002)]{kang02}
Kang, X., Jing, Y.\ P., Mo, H.\ J., \& B\"orner, G. 2002 MNRAS, 336, 892

\bibitem[Kauffmann et al.(1997)]{kauffmann97}
Kauffmann, G., Nusser, A., \& Steinmetz, M. 1997, MNRAS, 286, 795

\bibitem[Kravtsov et al.(2004)]{kravtsov04}
Kravtsov, A.\ V., Berlind, A.\ A., Wechsler, R.\ H., Klypin, A.\ A.,
Gottl\"ober, S., Allgood, B., Primack, J.\ R. 2004, ApJ, 609, 35

\bibitem[Kuhlen et al.(2004)]{kuhlen04}
Kuhlen, M., Strigari, L.\ E., Zentner, A.\ R., Bullock, J.\ S., Primack,
J.\ R., 2004, ApJ (submitted), astro-ph/0405143

\bibitem[Ma \& Fry(2000)]{ma00}
Ma, C. \& Fry, J.\ N. 2000, ApJ, 543, 503

\bibitem[Melott(1983)]{melott83}
Melott, A.\ L. 1983, ApJ, 264, 59

\bibitem[Mo et al.(2004)]{mo04}
Mo, H.\ J., Yang, X., \& van den Bosch, F.\ C. 2004, MNRAS, 349, 205

\bibitem[Navarro, Frenk, \& White(1997)]{nfw97}
Navarro, J.\ F., Frenk, C.\ S., \& White, S.\ D.\ M. 1997, ApJ, 490, 493

\bibitem[Park(1990)]{park90}
Park, C. 1990, MNRAS, 242, 59

\bibitem[Park et al.(1994)]{park94}
Park, C., Vogeley, M.\ S., Geller, M.\ J., \& Huchra, J.\ P. 1994, ApJ, 431, 569

\bibitem[Peacock \& Dodds(1994)]{pd94}
Peacock, J.\ A. \& Dodds, S. 1994, MNRAS, 267, 1020

\bibitem[Peacock et al.(2001)]{peacock01}
Peacock, J.\ A., \etal\ 2001, Nature, 410, 169

\bibitem[Peacock \& Smith(2001)]{peacocsmithk01}
Peacock, J.\ A., Smith, R.\ E. 2000, MNRAS, 318, 1144

\bibitem[Peebles(1976)]{peebles76}
Peebles, P.\ J.\ E. 1976, ApJ, 205, L109

\bibitem[Peebles(1979)]{peebles79}
Peebles, P.\ J.\ E. 1979, AJ, 84, 730

\bibitem[Percival et al.(2002)]{percival02}
Percival, W.\ J. et al. 2002, MNRAS, 337, 1068

\bibitem[Sargent \& Turner(1977)]{sargent77}
Sargent, W.\ L.\ W. \& Turner, E.\ L. 1977, ApJ, 212 L3

\bibitem[Scoccimarro et al.(2001)]{roman01}
Scoccimarro, R., Sheth, R.\ K., Hui, L., \& Jain, B. 2001, ApJ, 546, 20

\bibitem[Scoccimarro(2004)]{roman04}
Scoccimarro, R. 2004, Phys. Rev. D (in press), astro-ph/0407214

\bibitem[Seljak \& Zaldarriaga(1996)]{cmbfast}
Seljak U. \& Zaldarriaga M. 1996, ApJ, 469, 437

\bibitem[Seljak(2000)]{seljak00}
Seljak, U. 2000, MNRAS, 318, 203

\bibitem[Seljak(2001)]{seljak01}
Seljak, U. 2001, MNRAS, 325, 1359

\bibitem[Sheldon et al.(2004)]{sheldon04}
Sheldon, E., \etal\ 2004, AJ, 127, 2544

\bibitem[Sheth et al.(2001)]{Sheth01a}
Sheth, R.\ K., Hui, L., Diaferio, A., \& Scoccimarro, R.\ 2001, MNRAS, 325, 
1288

\bibitem[Sheth \& Diaferio(2001)]{Sheth01b}
Sheth, R.\ K. \& Diaferio, A. 2001, MNRAS, 322, 901

\bibitem[Spergel et al.(2003)]{spergel03}
Spergel, D.\ N., \etal\ 2003, ApJ, 148, 175

\bibitem[Springel, Yoshida, \& White(2001)]{gadget}
Springel, V., Yoshida, N., \& White, S.\ D.\ M. 2001 NewA, 6, 79

\bibitem[Strauss \& Willick(1995)]{strauss95}
Strauss, M.\ A. \& Willick, J.\ A. 1995 Phys. Rep., 261, 271

\bibitem[Strauss et al.(2002)]{strauss02} 
Strauss, M. A., et al.\ 2002, AJ, 124, 1810

\bibitem[Tegmark et al.(2004a)]{tegmark04a}
Tegmark, M., \etal\ 2004a, ApJ, 606, 702

\bibitem[Tegmark et al.(2004b)]{tegmark04b}
Tegmark, M., \etal\ 2004b, Phys. Rev. D, 69, 103501

\bibitem[Tinker et al.(2004)]{tinker04}
Tinker, J.\ L., Weinberg, D.\ H., Zheng, Z., \& Zehavi, I. 2004, ApJ (submitted), astro-ph/0411777

\bibitem[van den Bosch et al.(2003)]{vdb03}
van den Bosch, F.\ C., Yang, X., \& Mo, H.\ J. 2003, MNRAS, 345, 723

\bibitem[Verde et al.(2002)]{verde02}
Verde, L., \etal\ 2002, MNRAS, 335, 432

\bibitem[White(2001)]{white01a}
White, M. 2001, MNRAS, 321, 1

\bibitem[White et al.(2001)]{white01b}
White, M., Hernquist, L., \& Springel, V. 2001, ApJ, 550, L129

\bibitem[Yang et al.(2004)]{Yang04}
Yang, X., Mo, H.\ J., Jing, Y.\ P., van den Bosch, F.\ C., \& Chu, X.\ Q. 2004, MNRAS, 350, 1153

\bibitem[York et al.(2000)]{york00}
York, D., et al.\ 2000, AJ, 120, 1579

\bibitem[Yoshikawa et al.(2001)]{yoshikawa01}
Yoshikawa, K., Taruya, A., Jing, Y.\ P., \& Suto, Y. 2001, ApJ, 558, 520

\bibitem[Zehavi et al.(2004a)]{zehavi04a} 
Zehavi, I., et al.\ 2004a, ApJ, 608, 16

\bibitem[Zehavi et al.(2004b)]{zehavi04b}
Zehavi, I., et al.\ 2004b, ApJ (submitted), astro-ph/0408569

\bibitem[Zheng(2004)]{zheng04a}
Zheng, Z. 2004, ApJ, 610, 61

\bibitem[Zheng et al.(2004)]{zheng04} 
Zheng, Z., Berlind, A.\ A., Weinberg, D.\ H., Benson, A.\ J., Baugh, C.\ M.,
Cole, S., Dav\'e, R., Frenk, C.\ S., Katz, A., \& Lacey, C.\ G. 2004,
ApJ (submitted), astro-ph/0408569

\bibitem[Zheng et al.(2002)]{ztwb02}
Zheng, Z., Tinker, J.\ L., Weinberg, D.\ H., \& Berlind, A.\ A. 2002, ApJ, 575, 617

\end{thebibliography}
\end{document}